\documentclass[10pt,preprint]{aastex}

\usepackage{amssymb}
\usepackage{amsmath}
\usepackage{graphicx}

\newcommand{\pd}[2]{\frac{\partial #1}{\partial #2}}
\newcommand{\HALF}{\frac{1}{2}}
\renewcommand{\vec}[1]{\mathbf{#1}}
\newcommand{\tens}[1]{\mathbf{#1}}
\newcommand{\DS}{\displaystyle}
\newcommand{\vol}{{\mathcal V}}
\newcommand{\rhs}{\mathcal L}
\newcommand{\riemann}{{\cal R}}

\slugcomment{}
\shorttitle{PLUTO Code}
\shortauthors{Mignone et al.}

\begin{document}

\title{PLUTO: a Numerical Code for Computational Astrophysics}

\author{A. Mignone\altaffilmark{1,2},
        G. Bodo\altaffilmark{2}, 
        S. Massaglia\altaffilmark{1}, 
        T. Matsakos\altaffilmark{1}, 
        O. Tesileanu\altaffilmark{1}, 
        C. Zanni\altaffilmark{3} \and A. Ferrari\altaffilmark{1}}

\altaffiltext{1}{Dipartimento di Fisica Generale ``Amedeo Avogadro", 
              Universit\`a degli Studi di Torino, 
              via Pietro Giuria 1,
              10125 Torino, Italy}
\altaffiltext{2}{INAF/Osservatorio Astronomico di Torino,
                 Strada Osservatorio 20, 
                 10025 Pino Torinese, Italy}
\altaffiltext{3}{Laboratoire d'Astrophysique de l'Observatoire de Grenoble
                 414 Rue de la Piscine, 38041 Grenoble Cedex 09, France}

\begin{abstract}
   {We present a new numerical code, PLUTO, for the solution of hypersonic flows
    in 1, 2 and 3 spatial dimensions and different systems of coordinates.
    The code provides a multi-physics, multi-algorithm modular environment 
    particularly oriented towards the treatment of astrophysical flows in presence of
    discontinuities.
    Different hydrodynamic modules and algorithms may be independently selected 
    to properly describe Newtonian, relativistic, MHD or 
    relativistic MHD fluids.
    The modular structure exploits a general framework for integrating a 
    system of conservation laws, built on modern Godunov-type 
    shock-capturing schemes. 
    Although a plethora of numerical methods has been successfully 
    developed over the past two decades, the vast majority 
    shares a common discretization recipe, involving three general steps:
    a piecewise polynomial reconstruction followed by the solution of
    Riemann problems at zone interfaces and a final evolution stage.
    We have checked and validated the code against several benchmarks 
    available in literature. Test problems in 1, 2 and 3 dimensions
    are discussed.}
\end{abstract}


%%%%%%%%%%%%%%%%%%%%%%%%%%%%%%%%%%%%%%%%%%%%%%%%%%%%%%%%%%%%%%%%%%%%%%%%%%%
\section{Introduction}
%
%
%
%
%
%
%%%%%%%%%%%%%%%%%%%%%%%%%%%%%%%%%%%%%%%%%%%%%%%%%%%%%%%%%%%%%%%%%%%%%%%%%%%

Theoretical models based on direct numerical 
simulations have unveiled a new way toward a better 
comprehension of the rich and complex phenomenology associated with 
astrophysical plasmas.

Finite difference codes such as ZEUS \citep{SN92a, SN92b} or 
NIRVANA+ \citep{Z98} inaugurated this novel era and have been used by an increasingly 
large fraction of researchers nowadays.
However, as reported in \cite{F02}, the lack of upwinding
techniques and conservation properties have progressively 
moved scientist's attention toward more accurate and robust 
methods.
In this respect, the successful employment of the so-called high resolution 
shock-capturing (HRSC) schemes have revealed a mighty tool
to investigate fluid dynamics in non-linear regimes.
Some of the motivations behind their growing popularity
is the ability to model strongly supersonic flows while retaining 
robustness and stability.
The unfamiliar reader is refereed to the books of \cite{Toro97},
\cite{LeVeque98} and references therein for a more 
comprehensive overview.

Implementation of HRSC algorithms is based on a conservative formulation
of the fluid equations and proper upwinding requires an exact or 
approximate solution \citep{Roe86} to the Riemann problem, 
i.e., the decay of a discontinuity separating two constant states.
This approach dates back to the pioneering work of \cite{God59}, 
and it has now become the lead line in developing high resolution codes
examples of which include FLASH \citep[ for reactive hydrodynamics]{FLASH00},
the special relativistic hydro code GENESIS \citep{AIMM99},
the Versatile Advection Code \citep[VAC,][]{Toth96}
or the new NIRVANA \citep{Z04}.

Most HRSC algorithms are based on the so-called reconstruct-solve-average (RSA) 
strategy. In this approach volume averages are first reconstructed using piecewise
monotonic interpolants inside each computational cell. 
A Riemann problem is then solved at 
each interface with discontinuous left and right states and the solution
is finally evolved in time.
It turns out that this sequence of steps is quite general for many systems of 
conservation laws and, therefore, it provides a general framework under which we 
have developed a multi-physics, multi-algorithm high resolution code, PLUTO.
The code is particularly suitable for time-dependent, explicit computations of
highly supersonic flows in presence of strong discontinuities and it can be
employed under different regimes, i.e., classical, relativistic unmagnetized 
and magnetized flows.
The code is structured in a modular way allowing a new module to be
easily incorporated. This flexibility turns out to be quite important,
since many aspects of computational fluid dynamics are still in rapid development.
Besides, the advantage offered by a multi-physics, multi-solver code is also 
to supply the user with the most appropriate algorithms and, at the same time,
provide inter-scheme comparison for a better verification of the simulation
results.
PLUTO is entirely written in the C programming language and can run on either 
single processor or parallel machines, the latter functionality being implemented
through the Message Passing Interface (MPI) library.
The code has already been successfully employed in the context of stellar and 
extragalactic jets \citep{BRMMF03, MMB04, MMB05_ssr}, 
radiative shocks \citep{M05, MMB05}, 
accretion disks \citep{BCMTRF05, TBRMC06}, 
magneto-rotational instability, relativistic Kelvin-Helmholtz 
instability and so forth.

The paper is structured as follows: in \S\ref{sec:code_design} we give a description
of the code design; in \S\ref{sec:phys} we introduce the physics modules available
in the code; in \S\ref{sec:source} we give a short overview on source terms 
and non-hyperbolicity and in \S\ref{sec:verification} 
the code is validated against several standard benchmarks.

%%%%%%%%%%%%%%%%%%%%%%%%%%%%%%%%%%%%%%%%%%%%%%%%%%%%%%%%%%%%%%%%%%%%%%%%%%%
\section{Code Design}\label{sec:code_design}
%
%
%
%
%
%
%%%%%%%%%%%%%%%%%%%%%%%%%%%%%%%%%%%%%%%%%%%%%%%%%%%%%%%%%%%%%%%%%%%%%%%%%%%

%
\begin{figure}
\label{fig:flow}
\plotone{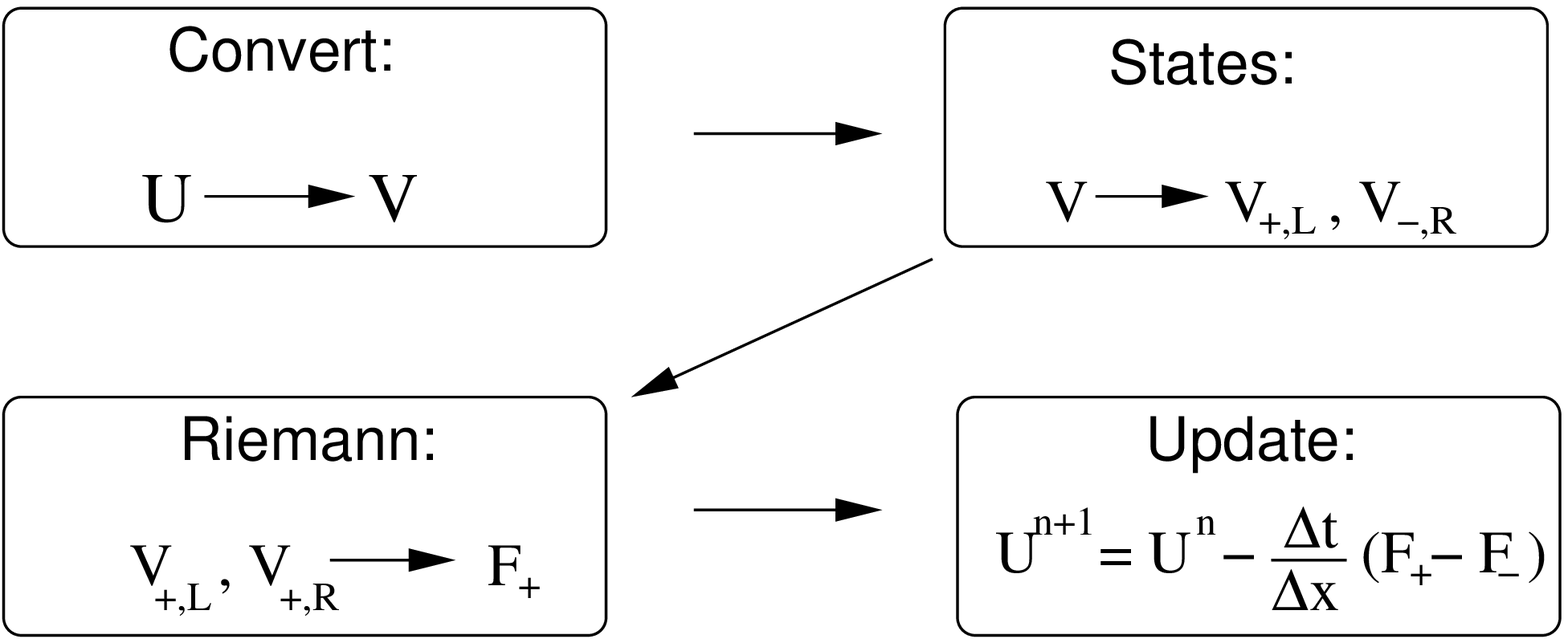}
\caption{Simplified flow diagram of the reconstruct-solve-average (RSA) 
         strategy: first, volume averages $\vec{U}$ are more conveniently
         mapped into primitive quantities $\vec{V}$. 
         Left and right states $\vec{V}_{+,\mathrm{L}}$ and 
         $\vec{V}_{-,\mathrm{R}}$ are constructed inside each zone 
         by suitable variable interpolation and/or extrapolation. 
         A Riemann problem is then solved between   
         $\vec{V}_{+,\mathrm{L}}$ and 
         $\vec{V}_{+,\mathrm{R}}$ to compute the numerical 
         flux function $\vec{F}_+$ at cell interfaces and the solution 
         is finally advanced in time.}
\end{figure}
PLUTO is designed to integrate a general system of conservation laws
which we write as
\begin{equation}\label{eq:CL}
  \pd{\vec{U}}{t} = - \nabla\cdot\tens{T}(\vec{U}) + \vec{S}(\vec{U}) \,.
\end{equation}
Here $\vec{U}$ denotes a state vector of conservative quantities,
$\tens{T}(\vec{U})$ is a rank-2 tensor, the rows of which are the fluxes
of each component of $\vec{U}$ and $\vec{S}(\vec{U})$
defines the source terms.
Additional source terms may implicitly arise when taking
the divergence of $\tens{T}(\vec{U})$ in a curvilinear 
system of coordinates.
An arbitrary number of advection equations may be added to the original
conservation law (\ref{eq:CL}).

Although the components of $\vec{U}$ are the primary variables being 
updated, fluxes are more conveniently computed using a different set
of physical quantities which we take as the primitive vector $\vec{V}$.
This choice is supported, moreover, by the fact that
interpolation on primitive variables enforces physical constraints such as 
pressure positivity and sub-luminal speeds in the case of relativistic flows.

Numerical integration of the conservation law (\ref{eq:CL}) is achieved
through shock-capturing schemes using the finite volume (FV) formalism
where \emph{volume averages} evolve in time.
Generally speaking, these methods are comprised 
of three steps: an interpolation routine followed by the solution of 
Riemann problems at zone edges and a final evolution stage.
In PLUTO, this sequence of steps provides the necessary 
infra-structure of the code, see the schematic diagram in Fig.
\ref{fig:flow}. 

At the higher level, the original system of equations is integrated by 
following this general sequence of steps, regardless any 
knowledge of the physics involved.
The explicit form of $\vec{U}$, $\vec{V}$, $\tens{T}(\vec{U})$ and $\vec{S}(\vec{U})$,
on the other hand, depends on the particular physical module selected.
Thus, at the lower level, a physical module collects the set of 
algorithms required to compute the terms involved in the 
discretization of the right hand side of Eq. (\ref{eq:CL}).
This set should provide one or more Riemann solver(s),
mapper routines for the conversion between primitive and conservative
variables, a flux routine giving the components
of $\tens{T}(\vec{U})$ in each direction, 
a source term function (if any) and a routine to compute the maximum
and minimum characteristic speeds of the Jacobian matrix.
Of course, additional features may be easily added by exploiting 
the independent modularity.

From the user's perspective, a particular configuration can be defined 
through a friendly interface entirely written in the Python scripting language.
The interface allows the user to specify all problem-dependent 
attributes and algorithms, such as number of dimensions, geometry, physics module, 
reconstruction method, time stepping integration and so forth.

%%%%%%%%%%%%%%%%%%%%%%%%%%%%%%%%%%%%%%%%%%%%%%%%%%%%%%%%%%%%%%%%%%%%%%%%%%%
\subsection{Notations}
%
%
%
%
%
%
%%%%%%%%%%%%%%%%%%%%%%%%%%%%%%%%%%%%%%%%%%%%%%%%%%%%%%%%%%%%%%%%%%%%%%%%%%%

\begin{deluxetable}{lccc}
\tablewidth{0pt} %.5\textwidth}
\tabletypesize{\footnotesize}
\tablecaption{Systems of coordinates (i.e. coordinates, volumes and areas) 
              adopted in PLUTO and their meaning.\label{tab:geom}}
\tablehead{
   \colhead{ } & \colhead{Cartesian} & \colhead{Polar} & \colhead{Spherical}
}
\startdata
 $x^1$          &     $x$      & $r$             &  $r$       \\
 $x^2$          &     $y$      & $\phi$          & $\theta$   \\
 $x^3$          &     $z$      & $z$             & $\phi$     \\
 $\Delta\vol^1$ & $\Delta x$   & $\Delta^2 r$    & $\Delta^3 r$ \\
 $\Delta\vol^2$ & $\Delta y$   & $ r\Delta\phi$  & $\Delta^3 r\Delta\mu$ \\
 $\Delta\vol^3$ & $\Delta z$   & $\Delta z$      & $\Delta^3r\Delta\mu\Delta\phi$ \\
 $A^1$          &     $1$      & $r_+ $          & $r^2_+$                        \\
 $A^2$          &     $1$      & $1$             & $\Delta^2 rs_+$                \\
 $A^3$          &     $1$      & $1$             & $\Delta^2r\Delta\theta$   
\enddata
\tablecomments{\footnotesize
         $r$ is the cylindrical or spherical radius, 
         $0\le\phi\le 2\pi$ is the azimuthal angle and $0\le\theta\le\pi$
         is the polar angle.
         Here $\Delta^n r= (r_+^n - r_-^n)/n$, $\Delta\mu=\cos\theta_- -\cos\theta_+$, 
         $s_+ = \sin\theta_+$, where $+$ or $-$ refer to the right or 
         left zone interface, respectively.
}
\end{deluxetable}
PLUTO employs logically rectangular grids in a generic system of 
orthogonal curvilinear coordinates $(x^1,x^2,x^3)$ (see Table \ref{tab:geom}).
Let $N_1$, $N_2$ and $N_3$ be the number of points in the three directions. 
The lower and upper coordinate bounds of zone $(i,j,k)$ are 
$(x^1_{i-\HALF}, x^2_{j-\HALF}, x^3_{k-\HALF})$ and 
$(x^1_{i+\HALF}, x^2_{j+\HALF}, x^3_{k+\HALF})$, respectively, where 
$1\le i\le N_1$, $1\le j\le N_2$, $1\le k\le N_3$.
The zone widths are then simply given by 
$\Delta x^1_i = x^1_{i+\HALF}-x^1_{i-\HALF}$,
$\Delta x^2_j = x^2_{j+\HALF}-x^2_{j-\HALF}$,
$\Delta x^3_k = x^3_{k+\HALF}-x^3_{k-\HALF}$.
The mesh spacing can be either uniform or 
(geometrical or logarithmic) stretched. 

Parallelization is achieved through domain decomposition: the global
domain is divided in sub-domains with an equal number of points and each 
of the sub-domains  is assigned to a processor. By default the sub-domains are 
created as maximally cubic, however the user, at run time can specify 
a different strategy for the creation of the sub-domains, imposing that
a given direction cannot be subdivided. In the code, parallelization is handled
through the MPI library. 
 
Grid adaptivity techniques are also being provided.
A one dimensional adaptive mesh refinement (AMR) module 
based on the \cite{BG89} method
is distributed with the code and extension to multiple spatial
dimensions is currently being developed using the 
CHOMBO library available at http://seesar.lbl.gov/ANAG/chombo/. 
However, since the main goal of this paper is to focus 
on the code modularity and its performance, AMR implementation 
will be described in future works.

In order to avoid formulas with cluttered notations we 
omit integer-valued subscripts when referring to 
three-dimensional quantities. Thus $\vec{V}_{i,j,k}$ becomes
simply $\vec{V}$. 
We introduce the standard two-point difference one-dimensional 
operator 
\begin{equation}\label{eq:rhs}
 \vec{\rhs}^{d}({\vec{V}}) = 
  -\frac{1}{\Delta\vol^d}\left(A^d_+\vec{F}^d_+ - A^d_-\vec{F}^d_-\right)  
  + \vec{S}^d  \,,
\end{equation}
where $d=1,2,3$ is a given direction, $A_\pm^d$ and $\Delta\vol^d$ are,
respectively, the cell's right ($+$) and left ($-$) interface areas 
and cell volume in that direction (see Table \ref{tab:geom}).
Here $\pm \equiv (i\pm\HALF,j,k),(i,j\pm\HALF,k),(i,j,k\pm\HALF)$ when $d=1,2,3$,
respectively.
Scalar components of the vectors appearing in  Eq. (\ref{eq:rhs}) will be denoted
with a pair of square brackets; e.g. $\rhs^{d}_{[m_\phi]}$ is the component 
of $\vec{\rhs}^{d}$ contributing to the $m_\phi$ equation.
Furthermore, since several of the building block algorithms are 
one-dimensional, we will also omit the superscript $d$ when unnecessary. 

The numerical flux functions $\vec{F}_\pm$ in Eq. (\ref{eq:rhs}) follow
the solution of 1-D Riemann problems at cell interfaces.
Since more than one Riemann solver may be available in each physics
module, we set, without loss of generality,
\begin{equation}
 \vec{F}_+ =
 \riemann\left(\vec{V}_{+,\mathrm{L}},\vec{V}_{+,\mathrm{R}}\right)  \,,
\end{equation}
where $\riemann$ is the Riemann solver being used and 
$\vec{V}_{+,\mathrm{L}}$, $\vec{V}_{+,\mathrm{R}}$ are suitable left and 
right states at the zone edges perpendicular to the $x$ direction.

%%%%%%%%%%%%%%%%%%%%%%%%%%%%%%%%%%%%%%%%%%%%%%%%%%%%%%%%%%%%%%%%%%%%%%%%%%%
\subsection{Reconstruction}
\label{sec:recon}
%
%
%
%
%
%
%%%%%%%%%%%%%%%%%%%%%%%%%%%%%%%%%%%%%%%%%%%%%%%%%%%%%%%%%%%%%%%%%%%%%%%%%%%

Interpolation routines are designed to reconstruct a piecewise 
polynomial approximations $\vec{\cal P}(\vec{x})$ to $\vec{V}$
inside each cell starting from its cell averages:
\begin{equation}\label{eq:interp}
  \vec{V}_{\pm,\mathrm{S}} = {\cal I}\left(\vec{\cal P}, \vec{V}\right) \,,
\end{equation}
where $\mathrm{S}=\mathrm{L}$ ($\mathrm{S}=\mathrm{R}$) 
at $\vec{x}=\vec{x}_+$ ($\vec{x}=\vec{x}_-$) .
Here ${\cal I}$ is an interpolation routine designated to 
provide left and right edge interpolated values inside each 
cell, that is, 
$\vec{V}_{+,\mathrm{L}} = \lim_{\vec{x}\to \vec{x}_+} \vec{\cal P}(\vec{x})$ and
$\vec{V}_{-,\mathrm{R}} = \lim_{\vec{x}\to \vec{x}_-} \vec{\cal P}(\vec{x})$,
where the ``L" and ``R" subscripts refer to the sides of the interface.

Reconstruction methods have to satisfy monotonicity constraints
in order to avoid spurious oscillations in proximity 
of discontinuities and steep gradients, see the books from 
\cite{Toro97}, \cite{LeVeque98} and references therein.
For second-order linear interpolants, for instance, one has
\begin{equation}
 \vec{V}_{\pm,\mathrm{S}} = \vec{V} \pm \frac{\Delta\tilde{\vec{V}}}{2} \,,
\end{equation}
with the slopes $\Delta\tilde{\vec{V}}$ computed following a limiting
procedure applied to primitive or characteristic variables. 
In the latter case one has
\begin{equation}
 \Delta\tilde{\vec{V}} = \sum_k \Delta\tilde{w}_k \vec{r}_k  
  \,, \quad
% \Delta\tilde{w}_k = \mathrm{limiter}(\vec{l}_k\cdot\Delta\vec{V}) \,,
 \Delta\tilde{w}_k = \mathrm{lim}\left(\Delta w_{k,+}, 
                                       \Delta w_{k,-}\right)\,,
\end{equation}
where $\vec{r}_k$ and $\vec{l}_k$ are, respectively, the right and 
left eigenvectors of the primitive form  of the equations, $\Delta w_{k,\pm} = 
\pm\vec{l}_k\cdot(\vec{V}_{i\pm 1} - \vec{V}_i)$ are forward ($+$) and
backward ($-$) derivatives and $k$ labels the $k$-th characteristic field. 
Different slope limiters ($\mathrm{lim}$) are characterized by distinct steepening properties
and can be independently assigned to each primitive variable or characteristic
field.

At the time of this writing, PLUTO allows to choose between 
either flat ($1^\mathrm{st}$ order in space), 
linear ($2^\mathrm{nd}$ order), third-order convex ENO \citep{dZB02}, 
parabolic reconstructions \citep[as in][]{MPB05}, or 
the 5th order finite difference WENO scheme of \cite{JS96} (only for the 
hydrodynamics equations).

%%%%%%%%%%%%%%%%%%%%%%%%%%%%%%%%%%%%%%%%%%%%%%%%%%%%%%%%%%%%%%%%%%%%%%%%%%%
\subsection{Riemann Solver}
%
%
%
%
%
%
%%%%%%%%%%%%%%%%%%%%%%%%%%%%%%%%%%%%%%%%%%%%%%%%%%%%%%%%%%%%%%%%%%%%%%%%%%%

As already anticipated, computation of the numerical flux 
function $\vec{F}_+$ at a zone edge ($x_+$) requires the solution 
$\vec{U}(x,t)$, for $t > t_0$, to the initial value problem
\begin{equation}\label{eq:riemann}
 \vec{U}(x,t_0) = \left\{\begin{array}{ccc}
   \vec{U}_{+,\mathrm{L}} & \quad \mathrm{if} \; & x < x_{+} \,, \\ \noalign{\medskip}
   \vec{U}_{+,\mathrm{R}} & \quad \mathrm{if} \; & x > x_{+} \,,
\end{array}\right.
\end{equation}
complemented with the one-dimensional evolutionary equation 
for $\vec{U}$.
Since this section deals with interface quantities, 
we shall omit, in this section only, the $+$ subscripts.

The exact solution to the Riemann problem (\ref{eq:riemann}) involves 
the decay of a set of non-linear waves and can be a rather cumbersome 
task to achieve. 
With the exception of few simple cases, existing Riemann solvers routinely 
involved in upwind schemes are based on different levels of 
approximation.

The Lax-Friedrichs Rusanov flux is robust, but also the most diffusive 
solver and is available for all modules. It computes the fluxes according
to:
\begin{equation}\label{eq:tvdlf}
 \vec{F} = \frac{1}{2}\Big[\vec{f}_{\mathrm{L}} + \vec{f}_{\mathrm{R}} - 
        \left|\lambda_{\max}\right|(\vec{U}_{\mathrm{R}} - \vec{U}_{\mathrm{L}})\Big] \,,
\end{equation}
where $\vec{f} = \hat{\vec{e}}^d\cdot\tens{T}(\vec{U})$ is the projection 
of the tensor flux on the 
$\hat{\vec{e}}^d = \left(\delta_{1d},\delta_{2d}, \delta_{3d}\right)$ unit vector,
$\delta_{ij}$ is Kronecker-Delta symbol 
and $|\lambda_{\max}|$ is the largest local signal velocity.

A somewhat different approximation comes from the HLL-family solvers:
\begin{equation}\label{eq:hll}
\vec{F}  = \left\{\begin{array}{cl}
 \vec{f}_{\mathrm{L}} & \;\mathrm{if} \;\; \lambda_1              > 0\,,   \\ \noalign{\medskip}
 \vec{f}_k            & \;\mathrm{if} \;\; \lambda_k\lambda_{k+1} < 0\;\; (k = 1,\dots,N-1)\,, \\ \noalign{\medskip}
 \vec{f}_{\mathrm{R}} & \;\mathrm{if} \;\; \lambda_N              < 0\,,   
 \end{array}\right.   
\end{equation}
where the solution to the Riemann problem is approximated by $N$ waves
with $\lambda_{k+1} > \lambda_k$ and $k = 1,\dots,N-1$ and separated 
by $N+1$ states.
The $\vec{f}_k$ are computed from a suitable ``parameterization" 
of the Rankine-Hugoniot jump conditions across each wave:
\begin{equation}
  \lambda_k\left(\vec{U}_{k+1} - \vec{U}_k\right) = 
  \vec{f}_{k+1} - \vec{f}_k  \,,
\end{equation}
with $\vec{U}_{N+1} = \vec{U}_{\mathrm{R}}$. 
The HLL and HLLC solvers can be consistently
derived following this recipe with $N=2$ and $N=3$, respectively, 
see \cite{TSS94}, \cite{BCLC97} 
for the hydro equations, \cite{Li05} for the MHD
equations, \cite{MB05,MB06} for the relativistic equations.
Similarly, we have also implemented the multi-state ($N=5$) HLLD solver of 
\citep{MK05} in the MHD module.
The HLL approach has some attractive features in that it guarantees 
positivity of pressure and, in the case of relativistic flows,
it preserves the condition $|\vec{v}| < 1$. 

Linearized Riemann solvers are more accurate since the averaging process 
inherent to Eq. (\ref{eq:tvdlf}) or (\ref{eq:hll}) is removed and all 
jumps are computed by linearization around some average state. 
The Roe solver computes the numerical fluxes according to:
\begin{equation}\label{eq:roe}
 \vec{F} = \frac{1}{2}\left[\vec{f}_{\mathrm{L}} + \vec{f}_{\mathrm{R}} 
     - \sum_k|\lambda_k| \vec{L}_k\cdot\left(
    \vec{U}_{\mathrm{L}} - \vec{U}_{\mathrm{R}}\right)\vec{R}_k\right] \,,
\end{equation}
where the rows (columns) of $\vec{L}$ ($\vec{R}$) are the left (right) eigenvectors 
of the Jacobian $\partial\vec{f}(\vec{U})/\partial\vec{U}$.

The Advection Upstream Splitting Method \cite[AUSM, originally proposed][]{LS93} 
provides an alternative and efficient 
upwinding strategy by splitting the flux into convective and 
pressure terms, respectively associated with linear and nonlinear fields:
\begin{equation}
 \vec{F} =  \vec{\Phi}v_n + \vec{p} \,,
\end{equation}
where $v_n$ is a suitably defined convective velocity and the $\vec{p}$ 
flux is governed by the acoustic speed.
The original AUSM scheme has been substantially improved in the work 
of \cite{Liou96, Liou06} and \cite{WL97}.

The most accurate approach consists in directly solving the whole set of
Rankine-Hugoniot jump conditions to find $\vec{V}^*$, from which 
the flux can be computed,
\begin{equation}
 \vec{F} = \vec{f}(\vec{V}^*) \,.
\end{equation}
However, this approach is computationally expensive 
since it generally involves the solution of highly 
nonlinear equations.

Different physics modules come with different sets of Riemann 
solvers, and additional methods of solution may be easily 
accommodated.
We warn that more accurate Riemann solvers (such linearized or
exact schemes) may introduce insufficient numerical dissipation 
for certain flow configurations. Sporadically, this could lead
to a number of numerical pathologies such as the carbuncle phenomena, 
odd-even coupling or lack of positivity-preserving properties. 
See the work by \cite{Quirk94} for a comprehensive review.

%%%%%%%%%%%%%%%%%%%%%%%%%%%%%%%%%%%%%%%%%%%%%%%%%%%%%%%%%%%%%%%%%%%%%%%%%%%
\subsection{Temporal Evolution}
%
%
%
%
%
%
%%%%%%%%%%%%%%%%%%%%%%%%%%%%%%%%%%%%%%%%%%%%%%%%%%%%%%%%%%%%%%%%%%%%%%%%%%%

Time marching algorithms provide a general (i.e. physics-independent) 
framework for the discretization of the left hand side
of Eq (\ref{eq:CL}). For example, in the simplest case
of forward Euler discretization one has
\begin{equation}
 \frac{\vec{U}^{n+1} - \vec{U}^n}{\Delta t} = \vec{\rhs}^n \,,
\end{equation}
where $\vec{\rhs}^n$ is the right hand side operator (\ref{eq:rhs}) or 
a sum of them for dimensionally split or unsplit schemes, respectively. 
The time step $\Delta t$ 
is limited by the Courant-Friedrichs-Lewy \citep[CFL,][]{CFL28} condition:
\begin{equation}\label{eq:CFL}
  \Delta t = C_\mathrm{a} \min_d\left(
            \frac{ \Delta l_{\min}^d }{ |\lambda_{\max}^d|}\right)  \,,
\end{equation}
with $\Delta l_{\min}^d$ and $\lambda_{\max}^d$ being, respectively,
the smallest cell length and largest signal velocity in the $d$ 
direction. $C_\mathrm{a}$ identifies the Courant number, restricted
by the conditions given in the last column of Table \ref{tab:nriem}.

PLUTO provides a number of time-marching schemes for the explicit 
numerical integration of the conservation law (\ref{eq:CL}).
We discriminate between 1) fully discrete, zone-edge extrapolated and 
2) semi-discrete methods.
Evolution in more than one dimension may be achieved by either operator
splitting \citep{Strang68} or fully multidimensional integration.

\begin{deluxetable}{lcccc}
\tabletypesize{\footnotesize}
\tablewidth{0pt}
\tablecaption{Number of Riemann problems per cell per time step required by different 
              time marching schemes in $N_d = 1, 2, 3$ dimensions.\label{tab:nriem}}
\tablehead{
   \colhead{Time Marching} & \colhead{1-D} & \colhead{2-D} & \colhead{3-D} & \colhead{$C_{\mathrm{a}}^{\max}$}
}
\startdata
  MH (S)      & 1   & 2     & 3    &  1  \\
  MH (U)      & -   & 4     & 12   &  1   \\
  ChTr (S)    & 1   & 2     & 3    &  1  \\
  ChTr (U)    & -   & 4     & 12   &  1   \\
  RK2 (S)     & 2   & 4     & 6    &  $1$\tablenotemark{\dagger}  \\
  RK2 (U)     & 3   & 6     & 9    &  $1/\sqrt{N_d}$  \\
  RK3 (S)     & 2   & 4     & 6    &  $1$  \\
  RK3 (U)     & 3   & 6     & 9    &  $1/\sqrt{N_d}$  
\enddata
\tablecomments{\footnotesize (S) or (U) stand for dimensionally split or unsplit algorithm, 
               MH and ChTr are the MUSCL-Hancock and Characteristic tracing 
               schemes, RK2 and RK3 are the $2^{\mathrm{nd}}$ and $3^{\mathrm{rd}}$ 
               order Runge-Kutta schemes. The rightmost column 
               gives the maximum allowed Courant number.}
\tablenotetext{\dagger}{High order interpolants (PPM, CENO3) may require a lower limit.}
\end{deluxetable}

%%%%%%%%%%%%%%%%%%%%%%%%%%%%%%%%%%%%%%%%%%%%%%%%%%%
\subsubsection{Zone-Edge Extrapolated Methods}
%
%
%%%%%%%%%%%%%%%%%%%%%%%%%%%%%%%%%%%%%%%%%%%%%%%%%%%

Zone-edge extrapolated methods achieve second order temporal accuracy 
by midpoint in time quadrature:
\begin{equation}
 \vec{U}^{n+1} = \vec{U}^n + \Delta t\vec{\rhs}^{n+\HALF} \,.
\end{equation}
and thus are based on a single-step.
For directional splitting methods one has 
$\vec{\rhs}^{n+\HALF}\equiv\vec{\rhs}^d\left(\vec{V}^{n+\HALF}\right)$, whereas 
$\vec{\rhs}^{n+\HALF} = \sum_d\vec{\rhs}^d\left(\vec{V}^{n+\HALF}\right)$ 
in the case of a dimensionally unsplit method. 
The input states for the Riemann solver are computed by Taylor
expansion of the primitive variables at half time step,
\begin{equation}\label{eq:states}
 \vec{V}^{n+\HALF}_{\pm,\mathrm{S}} = \vec{V}^n \pm \frac{1}{2}\left(\tens{I} - 
  \frac{\Delta t}{\Delta x}\tens{A}^{n}\right)\cdot\Delta\tilde{\vec{V}}^{n}  \,,
\end{equation}
where $\tens{A}^{n}$ is the matrix of the quasi-linear form of the 
equations and $\tens{I}$ is the identity matrix.
An alternative form  which does not require the primitive
form of the equations can be written in terms of 
conservative variables.
Both variants yield
the well-known MUSCL-Hancock scheme  
\citep[MH henceforth,][]{vL74, Toro97}.
When eigenvalues and eigenvectors are available, upwind limiting 
may be used to select only those characteristics which 
contribute to the effective left and right states. 
This approach is employed, for instance, in the original 
PPM advection scheme of \cite{CW84} and we will refer to as 
characteristic tracing (ChTr henceforth).
These methods require boundary conditions to be assigned at the 
beginning of the time step only.

The advantage offered by dimensionally split single-step algorithms is the 
lower computational cost (only one Riemann solver per cell per direction 
is required, see Tab \ref{tab:nriem}) and storage requirement.
%The dimensionally unsplit version of these schemes, however, adopts the corner 
An alternative, dimensionally unsplit version of these schemes, however, adopts the corner 
transport upwind (CTU) method of \cite{Colella90}, \cite{Saltzman94}
and is more expensive.
In this case, an extra correction term is needed in Eq. (\ref{eq:states}); 
in two dimensions, for example, the input states for the Riemann problem 
are modified to
\begin{equation}
  \tilde{\vec{U}}^{n+\HALF}_{\pm,\mathrm{S}} = \vec{U}^{n+\HALF}_{\pm,\mathrm{S}} + 
                   \frac{\Delta t}{2}\vec{\rhs}^{t,n+\HALF} \,,
\end{equation}
with $\vec{\rhs}^{t,n+\HALF}$ being the right-hand side operator corresponding to
the transverse direction.
As in the dimensionally split case, the CFL
stability condition is not affected by the dimensionality of the problem
(i.e. $C_a < 1$, see also Tab \ref{tab:nriem}).
Contrary to its dimensionally split version, on the other hand,
this method is computationally more expensive since 4 instead of 2
(in 2D) and 12 instead of 3 (in 3D) Riemann problems must be solved 
at each interface, see Table \ref{tab:nriem}.
%However, single-step time marching schemes offer simplicity of implementation
%on adaptive grids.

%%%%%%%%%%%%%%%%%%%%%%%%%%%%%%%%%%%%%%%%%%%%%%%%%%%
\subsubsection{Semi-Discrete Methods}
%
%
%%%%%%%%%%%%%%%%%%%%%%%%%%%%%%%%%%%%%%%%%%%%%%%%%%%

Semi-discrete methods are based on the classical method of lines, 
where the spatial discretization is considered separately from the 
temporal evolution which is left continuous in time.
In this framework Eq. (\ref{eq:CL}) is discretized as a regular ODE.
PLUTO implements the $2^{\mathrm{nd}}$ and $3^{\mathrm{rd}}$ 
order Total Variation Diminishing (TVD) Runge-Kutta schemes 
of \cite{GS96}. 
The second-order method (RK2) advances the system 
of conservation law (\ref{eq:CL}) as
\begin{equation}\label{eq:rk2_pred}
    \vec{U}^* = \vec{U}^n + \Delta t\vec{\rhs}^n   \,,
\end{equation}
\begin{equation}\label{eq:rk2_corr}
    \vec{U}^{n+1} = \frac{1}{2}\Big[\vec{U}^n + \vec{U}^* 
                 + \Delta t\vec{\rhs}^*\Big]  \,,
\end{equation}
with $\vec{\rhs}^n\equiv\vec{\rhs}^d\left(\vec{V}^n\right)$ or 
$\vec{\rhs}^n = \sum_d\vec{\rhs}^d\left(\vec{V}^n\right)$ in the 
case of a dimensionally split or unsplit method, respectively. 
The third-order Runge-Kutta method (RK3) may also be used, at the
cost of an additional step:
\begin{equation}
  \vec{U}^* = \vec{U}^n + \Delta t\vec{\rhs}^n  \,,
\end{equation}
\begin{equation}
  \vec{U}^{**} = \frac{1}{4}\Big[3\vec{U}^n + \vec{U}^* + \Delta t\vec{\rhs}^*\Big]  \,,
\end{equation}
\begin{equation}
  \vec{U}^{n+1} = \frac{1}{3}\Big[\vec{U}^n + 2\vec{U}^{**} + 2\Delta t\vec{\rhs}^{**}\Big]\,.
\end{equation}
For this class of methods, input states for the Riemann solver are given 
by the output of the interpolation routine, see \S\ref{sec:recon}.
Besides, boundary conditions must be assigned before each step.
A total of two and three Riemann problems per cell per
direction must be solved by the RK2 and RK3 marching scheme, respectively.
Furthermore, fully unsplit Runge-Kutta integrators require
a stronger time step limitation, see Table \ref{tab:nriem}.

%%%%%%%%%%%%%%%%%%%%%%%%%%%%%%%%%%%%%%%%%%%%%%%%%%%%%%%%%%%%%%%%%%%%%%%%%%%
\section{Physics Modules}\label{sec:phys}
%
%
%
%
%
%
%%%%%%%%%%%%%%%%%%%%%%%%%%%%%%%%%%%%%%%%%%%%%%%%%%%%%%%%%%%%%%%%%%%%%%%%%%%

PLUTO is distributed with four independent physics modules 
for the explicit numerical integration of the fluid equations under 
different regimes and conditions.
The hydrodynamics (HD), magnetohydrodynamics (MHD), relativistic (RHD), 
and relativistic MHD (RMHD) modules solve, respectively, the 
Euler equations of gas dynamics, the ideal/resistive MHD equations
the energy-momentum conservation laws of a special relativistic perfect gas, 
and the equations for a (special) relativistic magnetized ideal plasma.

In what follows, $\rho$, $p$, and $E$ will denote, 
respectively, the proper density, thermal pressure
and total energy density.
Vector fields such as 
$\vec{m} \equiv (m_1, m_2, m_3)^{\mathrm{T}}$, 
$\vec{v} \equiv (v_1, v_2, v_3)^{\mathrm{T}}$,
$\vec{B} \equiv (B_1, B_2, B_3)^{\mathrm{T}}$,
define the momentum density, velocity and magnetic field.
Finally, $\Gamma$ will be used to define the ratio of specific
heats for the ideal equation of state.

%%%%%%%%%%%%%%%%%%%%%%%%%%%%%%%%%%%%%%%%%%%%%%%%%%%
\subsection{The Hydrodynamics (HD) Module}
\label{sec:hd}
%
%
%%%%%%%%%%%%%%%%%%%%%%%%%%%%%%%%%%%%%%%%%%%%%%%%%%%

This module implements the equations of classical 
fluid dynamics with an ideal equation of state.
The conservative variables $\vec{U}$ and 
the flux tensor are:
\begin{equation}
 \vec{U} = \left(\begin{array}{c}
   \rho    \\ \noalign{\medskip}
   \vec{m}   \\ \noalign{\medskip} 
   E 
\end{array}\right)   \,, \quad
 \tens{T}\left(\vec{U}\right) = \left(\begin{array}{c}
   \rho\vec{v}           \\ \noalign{\medskip}
   \vec{m}\vec{v} + p\tens{I} \\ \noalign{\medskip}
   (E+p)\vec{v}   
\end{array}\right)^{\mathrm{T}}   \,,
\end{equation}
where $\vec{m} = \rho\vec{v}$ is the momentum density and 
$\tens{I}$ is the unit, $3\times 3$ tensor.
The total energy density $E$ is related to the gas pressure $p$ 
by the ideal gas closure:
\begin{equation}
  E = \frac{p}{\Gamma - 1} + \frac{|\vec{m}|^2}{2\rho} \,.
\end{equation}
The set of primitive variables $\vec{V}\equiv(\rho,\vec{v},p)^T$ is given by  
density, velocity $\vec{v}$ and thermal pressure $p$.

This module comes with a set of several Riemann solvers, 
including the nonlinear Riemann solver based on the 
two-shock approximations \citep{CW84,FLASH00}, 
the Roe solver \citep{Toro97}, the AUSM+ scheme 
\citep{Liou96}, the HLL \citep{ERM91}, 
HLLC \citep{TSS94} 
solvers and the Lax-Friedrichs solver \citep{Rus61}.

The HD module contains an implementation of the 
fast Eulerian transport algorithm for differentially
rotating disk \citep[FARGO,][]{Masset00} on polar
grids.
The FARGO scheme allows much larger time steps than the standard
integration where the Courant condition is traditionally limited by 
the fast orbital motion at the inner boundary.

%%%%%%%%%%%%%%%%%%%%%%%%%%%%%%%%%%%%%%%%%%%%%%%%%%%%%%%%%%
\subsection{The Magnetohydrodynamics (MHD) Module}
\label{sec:mhd}
%
%
%%%%%%%%%%%%%%%%%%%%%%%%%%%%%%%%%%%%%%%%%%%%%%%%%%%%%%%%%%

The MHD module deals with the equations of classical
ideal or resistive magnetohydrodynamics (MHD).
In the ideal case, $\vec{U}$ and $\tens{T}$ may be written as
\begin{equation}
 \vec{U} = \left(\begin{array}{c}
   \rho      \\ \noalign{\medskip}
   \vec{m}   \\ \noalign{\medskip} 
   \vec{B}   \\ \noalign{\medskip} 
   E 
\end{array}\right)   \,, \quad
 \tens{T}\left(\vec{U}\right) = \left(\begin{array}{c}
   \rho\vec{v}                        \\ \noalign{\medskip}
   \vec{m}\vec{v} - \vec{B}\vec{B} + p_{\mathrm{t}}\tens{I} \\ \noalign{\medskip}
   \vec{v}\vec{B} - \vec{B}\vec{v}    \\ \noalign{\medskip}
   (E + p_{\mathrm{t}})\vec{v} - (\vec{v}\cdot\vec{B})\vec{B}
\end{array}\right)^{\mathrm{T}}    \,,
\end{equation}
with $\vec{m} = \rho\vec{v}$ and $p_{\mathrm{t}} = p + |\vec{B}|^2/2$ 
being the total (thermal + magnetic) pressure, respectively.
The additional constraint $\nabla\cdot\vec{B} = 0$ complements 
the magnetic field evolution (see \S\ref{sec:divB}).
Resistivity is introduced by adding appropriate parabolic terms
to the induction and energy equations, see \S\ref{sec:diffusion}.
%More appropriately, the evolution of the magnetic field may be written 
%in terms of the induction equation, 
%
%\begin{equation}
% \pd{\vec{B}}{t} + \nabla\times\left( 
%  -\vec{v}\times\vec{B}
%+ \eta \nabla\times\vec{B} \right)  = 0   
%\end{equation}  
%
%where $\eta$ is the resistivity.

Available equations of state implemented are the ideal gas law,
\begin{equation}
  E = \frac{p}{\Gamma - 1} +
      \frac{1}{2}\left(\frac{|\vec{m}|^2}{\rho} + |\vec{B}|^2\right)\,.
\end{equation}
and the isothermal equation of state $p=c_s^2\rho$ where $c_s$ is the 
(constant) isothermal speed of sound.

The set of primitive variables is the same one used for the
HD module, with the addition of magnetic fields.
The user can choose among the following available Riemann solvers: 
the Roe solver of \cite{CG97}, the HLL \citep{J00}, HLLC \citep{Li05}, 
HLLD \citep{MK05} and the Lax-Friedrichs solvers.

%%%%%%%%%%%%%%%%%%%%%%%%%%%%%%%%%%%%%%%%%%%%%%%%%%%
\subsubsection{Solenoidal Constraint}
\label{sec:divB}
%
%
%%%%%%%%%%%%%%%%%%%%%%%%%%%%%%%%%%%%%%%%%%%%%%%%%%%

The solution to the MHD equations must fulfill
the solenoidal constraint, $\nabla\cdot\vec{B} = 0$, at all times.
Unfortunately it is well known that numerical scheme do 
not naturally preserve this condition unless special 
discretization techniques are used.
Among the variety of monopole control strategies 
proposed in literature \citep[for a review see][]{Toth00}, we 
have implemented 1) the eight wave formulation \citep{Pow94,PRL99}
and 2) the constrained transport (CT henceforth) 
of \cite{BS99}, and \cite{LdZ04}. 
The CT framework has been incorporated into the unsplit CTU 
integrator following the recent work by \cite{GS05}. A similar
approach is used by \cite{TFD06}.

In the first strategy, the magnetic field has a cell
centered representation and an additional source term
is added to the MHD equation.
The discretization of the source term is different depending
on the Riemann solver \citep[following][]{J00}, a feature which we found to greatly
improve robustness.
 
In the CT formulation, on the other hand, the induction equation is 
integrated directly using Stoke's theorem and the magnetic field
has a staggered collocation.

%%%%%%%%%%%%%%%%%%%%%%%%%%%%%%%%%%%%%%%%%%%%%%%%%%%%%%%%%%%%%%
\subsection{The Relativistic Hydrodynamics Module (RHD)}
\label{sec:rhd}
%
%
%%%%%%%%%%%%%%%%%%%%%%%%%%%%%%%%%%%%%%%%%%%%%%%%%%%%%%%%%%%%%%

This module deals with the motion of an ideal fluid
in special relativity.
The equations are given by particle number conservation and
energy-momentum conservation \citep{Lan_Lif59}.
Conservative variables and tensor flux are:
\begin{equation}
 \vec{U} = \left(\begin{array}{c}
   D      \\ \noalign{\medskip}
   \vec{m}   \\ \noalign{\medskip} 
   E 
\end{array}\right)      \,, \quad
 \tens{T}\left(\vec{U}\right) = \left(\begin{array}{c}
   D\vec{v}           \\ \noalign{\medskip}
   \vec{m}\vec{v} + p\tens{I} \\ \noalign{\medskip}
   \vec{m}
\end{array}\right)^{\mathrm{T}}   \,,
\end{equation}
where $D = \gamma\rho$ is the laboratory density and 
$\gamma = (1 - |\vec{v}|^2)^{-\HALF}$ is the Lorentz factor.
For convenience, velocities are normalized to the the speed of light.

The relation between conserved and primitive variables 
$\vec{V} \equiv (\rho, \vec{v}, p)^{\mathrm{T}}$ is more
involved than its classical counterpart. Specifically we have
\begin{equation}
  D = \DS \gamma \rho  \,,\quad
  \vec{m} = \DS \rho h \gamma^2 \vec{v} \,,\quad
  E = \DS \rho h \gamma^2 - p  \,,
\end{equation}
where the specific enthalpy $h\equiv h(p,\rho)$ depends on 
the equation of state.
The inverse map requires the solution of a nonlinear equation.
\cite{RCC06} address this topic for different choices of $h(p,\rho)$.

Two equations of state are available for this module: 
the constant-$\Gamma$ ideal gas law
\begin{equation}\label{eq:rhd_ideal}
 h = 1 + \frac{\Gamma}{\Gamma - 1}\Theta\,,
\end{equation}
with $\Theta = p/\rho$ and the equation of state
(TM henceforth) already introduced in \cite{MPB05}:
\begin{equation}\label{eq:rhd_tm}
 h = \frac{5}{2}\Theta + \sqrt{\frac{9}{4}\Theta^2 + 1} \,,
\end{equation}
which satisfies Taub's inequality \citep{Taub48} and approximates 
(within $\la 4\%$) the single-specie relativistic 
perfect gas \citep[for a thorough discussion, see][ and references
therein]{MPB05, RCC06}. 

The two shock nonlinear Riemann solver described in \cite{MPB05}
has been incorporated into the set of available Riemann
solvers, together with the recently developed HLLC 
algorithm by \cite{MB05}.
The HLL and Lax-Friedrichs Riemann solvers have also been
coded.

%%%%%%%%%%%%%%%%%%%%%%%%%%%%%%%%%%%%%%%%%%%%%%%%%%%%%%%%%%%%%%%%%%%%%%%
\subsection{The Relativistic Magnetohydrodynamics Module (RMHD)}
\label{sec:rmhd}
%
%
%%%%%%%%%%%%%%%%%%%%%%%%%%%%%%%%%%%%%%%%%%%%%%%%%%%%%%%%%%%%%%%%%%%%%%%

The motion of an ideal relativistic magnetized fluid is
described by conservation of mass, energy-momentum,
and by Maxwell's equations, see, for example, \cite{AP87, Anile89}.
PLUTO implements the equation of special relativistic
magnetohydrodynamics where the vector of conservative variables 
and respective fluxes can be cast as
\begin{equation}
\vec{U}= \left(\begin{array}{c}
    D    \\ \noalign{\medskip}
   \vec{m} \\ \noalign{\medskip}
   \vec{B} \\ \noalign{\medskip}
   E   \\ \noalign{\medskip}
\end{array}\right)     \,, \quad
 \tens{T}\left(\vec{U}\right) = \left(\begin{array}{c}
   D\vec{v}    \\ \noalign{\medskip}
   w_{\mathrm{t}}\gamma^2\vec{v}\vec{v} - \vec{b}\vec{b} + \tens{I}p_{\mathrm{t}} \\ \noalign{\medskip}
   \vec{v}\vec{B} - \vec{B}\vec{v}    \\ \noalign{\medskip}
  \vec{m}
\end{array}\right)^{\mathrm{T}}   \,,
\end{equation}
where $w_{\mathrm{t}} = \rho h + b^2_{\mathrm{m}}$,
$b^2_{\mathrm{m}} = |\vec{B}|^2/\gamma^2 + (\vec{v}\cdot\vec{B})^2$, 
$\vec{b} = \vec{B}/\gamma + \gamma (\vec{v}\cdot\vec{B})\vec{v}$ and
$p_{\mathrm{t}} = p + b_{\mathrm{m}}^2/2$ is the total (thermal + magnetic) pressure.

The components of $\vec{U}$ are related to the primitive variables 
$\vec{V} \equiv (\rho, \vec{v}, p, \vec{B})$ through
\begin{eqnarray}
   D   & = & \rho\gamma   \;,  \label{eq:cons_var_D}\\ \noalign{\medskip}
\vec{m}& = & (\rho h\gamma^2 + |\vec{B}|^2)\vec{v} - (\vec{v}\cdot\vec{B})\vec{B}  \; ,
          \label{eq:cons_var_m}\\ \noalign{\medskip}
 E   & = & \DS \rho h\gamma^2  - p
            + \frac{|\vec{B}|^2}{2} + \frac{|\vec{v}|^2|\vec{B}|^2 
           - (\vec{v}\cdot\vec{B})^2}{2}
             \label{eq:cons_var_E}\,.
\end{eqnarray}
The inverse map is highly nonlinear and different methods of solution have been
proposed when a constant $\Gamma$ law is adopted \citep[for a recent review, see][]{NGmCdZ06}. 
In PLUTO, we follow the approach described in \cite{MB06}
which was recently generalized to include the TM equation of state
described in \S\ref{sec:rhd}. The interested reader should also refer to 
the work by \cite{MK07}. 

The RMHD module may be used with the Lax-Friedrichs, HLL or 
the recently developed HLLC Riemann solver, see \cite{MB06}.
Since the induction equation takes the same form as in the classical 
case, the divergence of the magnetic field can be kept under control by any 
of the methods already introduced in the MHD module, see \S\ref{sec:divB}.

%%%%%%%%%%%%%%%%%%%%%%%%%%%%%%%%%%%%%%%%%%%%%%%%%%%%%%%%%%%%%%%%%%%%%%%%%%%
\section{Source Terms and non-Hyperbolicity}
\label{sec:source}
%
%
%
%
%
%
%%%%%%%%%%%%%%%%%%%%%%%%%%%%%%%%%%%%%%%%%%%%%%%%%%%%%%%%%%%%%%%%%%%%%%%%%%%

In addition to the homogeneous terms previously described, 
a number of additional features may be added in the code.

Local source terms are functions of the variables themselves but not of
their derivatives and are included either during the advection step
or via operator splitting.
Examples include the centripetal and Coriolis terms implicitly arising 
when working in polar or spherical coordinates,
external forces (e.g. gravity) or optically thin radiative losses.

Non-ideal effects such as viscosity, resistivity and thermal 
conduction, on the other hand, introduce parabolic corrections to the equations
and involve the solution of diffusion equations. 

%%%%%%%%%%%%%%%%%%%%%%%%%%%%%%%%%%%%%%%%%%%%%%%%%%%%%%%%%%%%%%%%%
\subsection{Geometrical Source Terms}
%
%
%
%%%%%%%%%%%%%%%%%%%%%%%%%%%%%%%%%%%%%%%%%%%%%%%%%%%%%%%%%%%%%%%%%

The divergence of a rank-2 tensor in curvilinear coordinates 
breaks down the homogeneous properties of Eq. (\ref{eq:CL}).
This loss of conservation comes from the additional source terms 
inevitably introduced by those unit vectors that do not have fixed 
orientation in space \cite[see, for instance, the Appendix in][]{MPB05}.
In the case of symmetric or anti-symmetric tensors, however, 
some of the source terms can be eliminated by properly re-arranging
the derivatives. This ensures conservation of the angular momenta rather
than the linear ones.

For all physics modules previously introduced, for example, the 
components of the flux tensor in the momentum equation 
form a $3\times 3$ rank 2 symmetric tensor $M_{ij} + p\delta_{ij}= 
M_{ji} + p\delta_{ji}$, where $p$ is the isotropic pressure term.
However, since $\nabla\cdot(\tens{I}p) \equiv \nabla p$, the differencing terms 
involving pressure are separately treated as gradient components 
and never contribute to the source terms. 

The symmetric character of $\tens{M}$ leads to further simplifications
in polar coordinates, since the $\phi$-component 
of the divergence of $\tens{M}$ may be written as
\begin{equation}
 (\nabla\cdot\tens{M})_\phi = \frac{1}{r^2}\pd{\left(r^2M_{r \phi}\right)}{r} + 
    \frac{1}{r}\pd{M_{\phi\phi}}{\phi} + 
               \pd{M_{z\phi}}{z} \,.
\end{equation}
This leads to the so-called ``angular momentum-conserving form" where
the radial contribution to $m_\phi$ is computed by the new operator
\begin{equation}
 \rhs^{r}_{[m_\phi]} = 
  -\frac{1}{r\Delta\vol^r}\left(r^2_+ F^r_{+,[m_\phi]} - r^2_-F^r_{-,[m_\phi]}\right)\,,
\end{equation}
thus leaving the radial component of momentum as the only non-homogeneous 
equation, with $S_{[m_r]} = M_{\phi\phi}/r$.
Similar considerations lead to the angular momentum conserving 
form in spherical geometry.

Geometrical source terms are integrated explicitly during the advection 
step by adding their contribution to the right hand side operator (\ref{eq:rhs}).
In the semi-discrete approach, this is straightforward.  
For edge-extrapolated methods, this requires 1) augmenting (\ref{eq:states}) with the 
cell-center source term for a half step and then 2) averaging the resulting time 
centered left and right edge values to compute the $\vec{S}$ for the final conservative step.
%computation of cell-centered time-centered 
%values by averaging $\vec{V}^{n+1/2}_{\pm,S}$ available from (\ref{eq:states}).

%%%%%%%%%%%%%%%%%%%%%%%%%%%%%%%%%%%%%%%%%%%%%%%%%%%
\subsection{Optically thin cooling}
%
%
%%%%%%%%%%%%%%%%%%%%%%%%%%%%%%%%%%%%%%%%%%%%%%%%%%%

For many astrophysical applications, radiative processes 
may become important during the evolution.
This is particularly true when the cooling timescale 
becomes comparable to, or smaller than the typical timescale for the dynamical 
evolution.

Optically thin radiative losses are treated as local source terms 
depending on the temperature, density and an arbitrary number
of ions of different elements (e.g. H, He, C, N, O, and so on).
The ionization fractions are advected with the fluid and 
are subject to collisional ionization, recombination 
and charge exchange processes.

Currently, we have implemented a cooling module \citep{TMM07} 
similar to the one coded by \cite{RML97}, extending the
temperature range of applicability ($2000 \la T \la 2 \cdot 10^5$ K) 
by using an increased number of ion species. 
This module employs $28$ ions and is valid for densities 
up to $10^5$ cm$^{-3}$.

Operator splitting is employed to evolve the chemical network in time.
A $2^{\rm nd}$ order fully implicit scheme is adopted in regions 
of rapid variations, whereas a second order explicit integrator
is used otherwise. A comprehensive description is outside the goal 
of this work and will be available in future works. 

%%%%%%%%%%%%%%%%%%%%%%%%%%%%%%%%%%%%%%%%%%%%%%%%%%%
\subsection{Treatment of Parabolic Terms}
\label{sec:diffusion}
%
%
%%%%%%%%%%%%%%%%%%%%%%%%%%%%%%%%%%%%%%%%%%%%%%%%%%%

Parabolic terms introduce second-order spatial derivatives
and their treatment requires the solution of diffusion equations.
Typical examples include electric resistivity,
\begin{eqnarray}\label{eq:resmhd}
 \pd{\vec{B}}{t} + \nabla\times\left( 
  -\vec{v}\times\vec{B}
+ \eta\vec{J} \right)  &=& 0  \,,  \\
 \pd{E}{t} + \nabla\cdot\left[(E + p_t)\vec{v} 
  - (\vec{v}\cdot\vec{B})\vec{B} 
  + \eta \vec{J}\times\vec{B}\right] &=&  0      \,,
\end{eqnarray}
where $\vec{J}=\nabla\times\vec{B}$ is the current and $\eta$ is the resistivity; 
or thermal conduction,
\begin{equation}
 \pd{E}{t} + 
 \nabla\cdot\left(\vec{F}^{\mathrm{adv}} - \kappa\nabla T\right) = 0\,,
\end{equation} 
where $\vec{F}^{\mathrm{adv}}$ is the energy advection flux
and $\kappa$ is the thermal conductivity coefficient.
These terms may be included into the original conservation 
law with the further time step limitation $\Delta t = 
\min(\Delta t^{\mathrm{ad}}, \Delta t^{\mathrm{par}})$, 
where $\Delta t^{\mathrm{ad}}$ is the advective time 
step (Eq. \ref{eq:CFL}) and 
\begin{equation}\label{eq:diff_timestep}
 \Delta t^{\mathrm{par}} < 0.5 \min_{d=1,2,3}\left[\frac{\left(
     \Delta x^d\right)^2}{\max(\sigma)}\right]\,.
\end{equation}
is the diffusion step.
Here $\sigma$ characterizes the physical process, e.g., $\eta$ 
or  $\kappa$.

For advection-dominated problems (i.e. 
$\Delta t^{\mathrm{ad}} \ga \Delta t^{\mathrm{par}}$), 
Eq. (\ref{eq:diff_timestep}) does not
pose severe restrictions and diffusion terms can be treated 
explicitly. 
We achieve this by adding centered in space, second-order finite difference 
approximations to the right hand side operator $\vec{\rhs}^d$, see Eq. (\ref{eq:rhs}).

However, for diffusion-dominated problems and/or increasingly high 
resolution simulations, $\Delta t^{\mathrm{par}}$ will eventually 
drop below the typical advection scale.
In such situations the explicit integration can be considerably 
accelerated using the super time stepping (STS) method 
\citep{AAG96,OD06}.
In STS, diffusion terms are included via operator splitting, 
and the solution vector is evolved over a super time step $\Delta T$ 
consisting of $N$ smaller sub-steps, the stability of which 
is closely related to the properties of Chebyshev polynomials. 
It can be proved \citep{AAG96} that 
\begin{equation}\label{eq:superstep}
\Delta T  \, \mathop{\to} \, N^2\Delta t^{\mathrm{par}} \,,
\end{equation}
which makes STS almost $N$ times faster than the standard explicit scheme.
Thus, if $\Delta T$ is taken to be the advective time step, STS will require 
(approximately) $\sqrt{\Delta t^{\mathrm{ad}}/\Delta t^{\mathrm{par}}}$ 
iterations rather than $\Delta t^{\mathrm{ad}}/\Delta t^{\mathrm{par}}$, 
typical of a normal sub-cycling explicit time stepping.
This approach offers dramatic ease of implementation over implicit schemes.

Resistivity has been extensively tested through the direct comparison with 
analytical solutions of linear diffusion problems of magnetic field.
Our results confirm the efficiency and accuracy of the STS approach already 
highlighted by previous investigators.
This has encouraged further experiments towards nonlinear problems 
(e.g. Spitzer-like conductivity), where preliminary results suggest 
that the STS methodology may be succesfully applied. 
This issue will be presented in future works.

%The tests were done in 2.5D and 3D in all coordinate systems using both the 
%8-wave formulation and the constrained transport.

%%%%%%%%%%%%%%%%%%%%%%%%%%%%%%%%%%%%%%%%%%%%%%%%%%%%%%%%%%%%%%%%%%%%%%%%%%
\section{Code Verification}\label{sec:verification}
%
%
%
%
%
%
%%%%%%%%%%%%%%%%%%%%%%%%%%%%%%%%%%%%%%%%%%%%%%%%%%%%%%%%%%%%%%%%%%%%%%%%%%%

\begin{deluxetable}{lcccccccc}
\tabletypesize{\scriptsize}
\tablewidth{\textwidth}
\tablecaption{Numerical schemes adopted in the different test problems described 
              in the text. \label{tab:tests}}
\tablehead{
\colhead{Test} & \colhead{Module} & \colhead{Case} & \colhead{Geometry} & \colhead{Dim} &
\colhead{Interpolation} & \colhead{Solver} & \colhead{Time Stepping} & \colhead{$C_a$}
}
\startdata
Double Mach Reflection(\S\ref{sec:hd_dmr})          & HD   & (a) & Cartesian   &  2  & Parabolic  & HLLC      & RK3  (S)  &  0.8   \\ 
                                                    &      & (b) &             &     & CENO3      & HLLC      & RK3  (S)  &  0.8    \\ 
                                                    &      & (c) &             &     & WENO5      & Roe F-S   & RK3  (S)  &  0.8    \\ 
                                                    &      & (d) &             &     & Parabolic  & Two-Shock & ChTr (S)  &  0.8    \\ 
                                                    &      & (e) &             &     & Linear     & Roe       & MH  (U)  &  0.8    \\ 
                                                    &      & (f) &             &     & Linear     & HLL       & MH  (U)  &  0.8    \\ \tableline
Under-expanded Jet (\S\ref{sec:hd_jet})             & HD   & (a) & Cylindrical &  2  & Linear     & AUSM+     & MH  (U)  &  0.4    \\ 
                                                    &      & (b) &             &     & CENO3      & HLL       & RK3  (S)  &  0.4    \\ \tableline
Rotated Shock Tube (\S\ref{sec:mhd_st2.5})          & MHD  & (a) & Cartesian   & 2.5 & Linear     & Roe       & RK2  (U)  &  0.4    \\
                                                    &      & (b) &             &     & Linear     & HLLD      & RK2  (U)  &  0.4    \\
                                                    &      & (c) &             &     & Linear     & HLL       & RK2  (U)  &  0.4    \\ \tableline
Fast Rotor (\S\ref{sec:mhd_rotor})                  & MHD  & (a) & Cartesian   &  2  & Parabolic  & HLLC      & ChTr (U)  &  0.6   \\  
                                                    &      & (b) & Polar       &     & Linear     & Roe       & RK3  (U)  &  0.4    \\ \tableline    
%Shock-Cloud Interaction(\S\ref{sec:mhd_sc3d})       & MHD  & (a) & Cartesian   &  3  & Linear     & Roe       & MH  (S)  &  0.6   \\ 
%                                                    &      & (b) &             &     & Linear     & Roe       & ChTr (S)  &  0.4    \\ \tableline
Magnetized Torus (\S\ref{sec:mhd_torus})            & MHD  & (a) & Spherical   & 2.5 & Linear     & HLLD      & RK2  (U)  &  0.4   \\ 
                                                    &      & (b) & Spherical   &     & Parabolic  & HLL       & RK3  (U)  &  0.4    \\
                                                    &      & (c) & Cylindrical &     & Linear     & HLLD      & RK2  (U)  &  0.4    \\
                                                    &      & (d) & Cylindrical &     & Parabolic  & HLL       & RK3  (U)  &  0.4    \\ \tableline
Spherical Shock Tube (\S\ref{sec:mhd_torus})        & RHD  & (a) & Spherical   &  1  & Parabolic  & Two-Shock & ChTr(S)  &  0.8   \\ 
                                                    &      & (b) & Spherical   &  1  & Linear     & HLL       & MH  (S)  &  0.8    \\
                                                    &      & (c) & Cartesian   &  3  & Linear     & HLLC      & ChTr(S)  &  0.8    \\ \tableline
Magnetized Blast Wave (\S\ref{sec:rmhd_blast3d})    & RMHD & (a) & Cartesian   &  3  & Linear     & HLL       & RK2  (U)  &  0.2    \\ 
                                                    &      & (b) & Cylindrical &  2  & Linear     & HLL       & MH  (U)  &  0.4    
\enddata
\tablecomments{\footnotesize Here ChTr stands for the characteristic tracing, 
              RK2 and RK3 for the $2^{\mathrm{nd}}$ and $3^{\mathrm{rd}}$ 
              Runge-Kutta time stepping, whereas MH is the MUSCL-Hancock time marching 
              scheme. Split or unsplit schemes are denoted with (S) or (U).
              The rightmost column gives the CFL number $C_\textrm{a}$.}
\end{deluxetable}

PLUTO has been validated against several benchmarks typically adopted for
other numerical schemes.
Several algorithms for relativistic flows presented in \cite{MPB05}
and \cite{MB05, MB06} are now part of the code and the related
verification problems will not be repeated here.
In what follows, we propose additional tests aimed to check the code performances
on problems of different nature, geometry and dimension.
Computational details for each test are given in Table \ref{tab:tests}.

%%%%%%%%%%%%%%%%%%%%%%%%%%%%%%%%%%%%%%%%%%%%%%%%%%%%%%%%%%%%
\subsection{Double Mach Reflection of a Strong Shock}
\label{sec:hd_dmr}
%
%%%%%%%%%%%%%%%%%%%%%%%%%%%%%%%%%%%%%%%%%%%%%%%%%%%%%%%%%%%%

\begin{figure*}
 \plotone{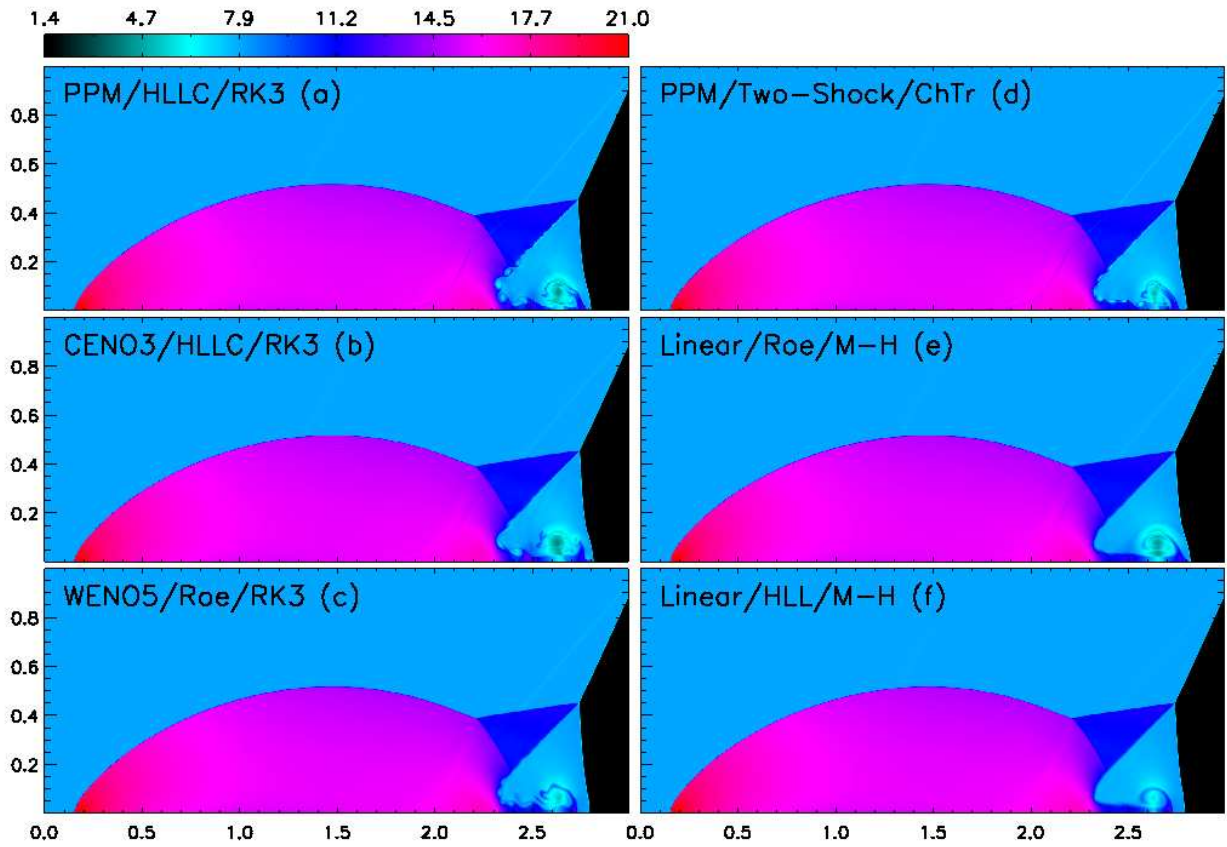}
 \caption{\footnotesize Density maps for the double Mach reflection test at $t=0.2$. 
          Each panel shows results obtained with the different combinations 
          of schemes listed in table \ref{tab:tests}. 
          The mesh size ($1/\Delta x = 1/\Delta y = 480$) and Courant number
          ($C_a=0.8$) are the same for all cases. For the sake of clarity only
          the region $[0,3]\times[0,1]$ is shown.}
 \label{fig:hd_dmr}
\end{figure*}
The initial condition for this test problem \citep{WC84}
consists in a planar shock front making an angle $\alpha = \pi/3$  
with a reflecting wall, taken to be the $x$ axis:
\begin{equation}
\Big(\rho, v_{\mathrm{x}}, v_{\mathrm{y}}, p \Big) = 
\left\{\begin{array}{cl}
  \Big(1.4, 0, 0 , 1 \Big)           & \mathrm{for} \quad x > x_{\mathrm{s}}(0) \\ \noalign{\medskip}
  \Big(8  , 8.25, -8.25, 116.5 \Big) & \mathrm{otherwise}  \,,  \\ \noalign{\medskip}
\end{array}\right.
\end{equation}
where $x_{\mathrm{s}}(t) = (10t/\sin\alpha + 1/6 + y/\tan\alpha)$ is the
shock coordinate. 
The ideal equation of state with $\Gamma = 1.4$ is used throughout
the calculation.
The computational domain is the rectangle $[0,4]\times[0,1]$ covered with 
a uniform grid with $1/\Delta x = 1/\Delta y = 480$. 
Outflow boundary conditions apply at $x=4$ and reflective conditions
are imposed at the bottom boundary $y=0$ for $x > 1/6$.
Fluid variables are kept constant to their initial values 
at $y=0$ for  $x < 1/6$ and at the leftmost boundary $x=0$.
At the top boundary $y=1$ the exact motion of the shock 
is prescribed.
We carry out integration until $t=0.4$ using the six different 
combinations of algorithms described in Table \ref{tab:tests}.
Panels in the left column of Fig. \ref{fig:hd_dmr} show,
from top to bottom, the results obtained with high order ($>2$) 
interpolants: piecewise parabolic reconstruction (a), $3^{\rm rd}$ 
order central ENO (b) and the WENO scheme of \cite{JS96} (c).
Panels on the right show the original
PPM scheme of \cite{CW84} (case d, top), the $2^{\rm nd}$ order 
unsplit Muscl-Hancock with a Roe solver (case e, middle) and with the 
HLL solver (case f, bottom).

After the reflection, a complicated flow structure develops with
two curved reflected shocks propagating at directions almost orthogonal 
to each other and a tangential discontinuity separating them. 
At the wall, a pressure gradient sets up a denser fluid jet 
propagating along the wall. 
Kelvin-Helmholtz instability patterns may be identified with the ``rolls" 
developing at the slip line. This feature is visible only for some of the 
selected schemes.
Indeed, the use of high-order interpolants, such as PPM (a,d) or WENO(c),
and the ability of the Riemann solvers to resolve contact 
and shear waves result in smaller numerical viscosity when compared to a 
second-order slope-limited reconstruction and/or a more approximate 
Riemann solver. In this respect, case a) shows the smallest amount of 
dissipation, whereas the HLL solver (f, bottom right) has the largest 
numerical diffusion.  

In terms of CPU time, we found that case a) to f) followed the ratios 
$1.82:1.55:8.52:0.94:1.24:1$, that is, with the $5^{\rm th}$ order WENO 
schemes being by far the most expensive (more than a factor of $8$ compared
to the simple HLL) and the original PPM (split) scheme being the
fastest. In doing the comparison, however, one has to bear in mind that the 
amount of computing time depends, in the first place, by the number of 
Riemann problems solved in each zone at each time step, as shown in Table \ref{tab:nriem}.
This number is 6 for cases a), b) and c), but $4$ for
case e) and f) (which are dimensionally unsplit) and only $2$ for the original dimensionally
split PPM scheme, case d).
Thus one can safely conclude, comparing case e) and f), that the Roe solver
is by a factor $\sim 1/4$ slower than HLL or, by comparing a), b) and c), that 
the CENO and PPM are considerably faster than WENO, although the combinations
of algorithms in a) better resolves small scale structures.

Finally, we note that computations carried with more accurate schemes (with the exception of 
the $5^{\rm th}$ order WENO) exhibit a slight tendency to form a spurious ``kinked" 
Mach stem on the $x$ axis. This feature is engendered by a numerical flaw \citep{Quirk94}
caused by insufficient numerical dissipation and it becomes particularly dramatic 
when the resolution is further increased.

%%%%%%%%%%%%%%%%%%%%%%%%%%%%%%%%%%%%%%%%%%%%%%%%%
\subsection{Under-expanded Jet}
\label{sec:hd_jet}
%
%
%
%%%%%%%%%%%%%%%%%%%%%%%%%%%%%%%%%%%%%%%%%%%%%%%%%

\begin{figure}
 \epsscale{1.0}
 \plotone{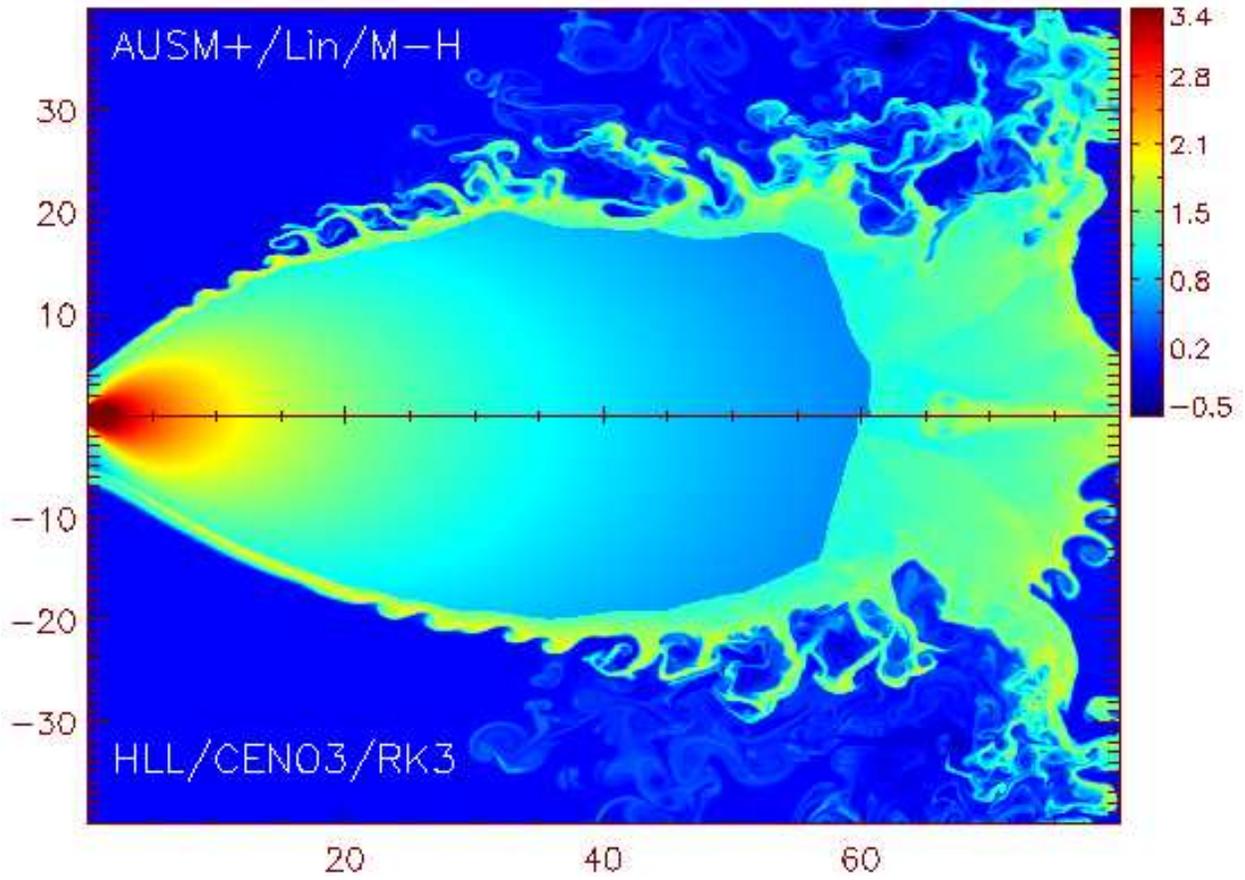}
 \caption{\footnotesize Density logarithms for the under-expanded jet at $t=240$,
          for the linear Muscl-Hancock CTU with the AUSM+ solver (top) and
          the RK3 with HLL solver and third-order CENO interpolation (bottom).}
 \label{fig:hdjet_a}
\end{figure}

\begin{figure}
 \plotone{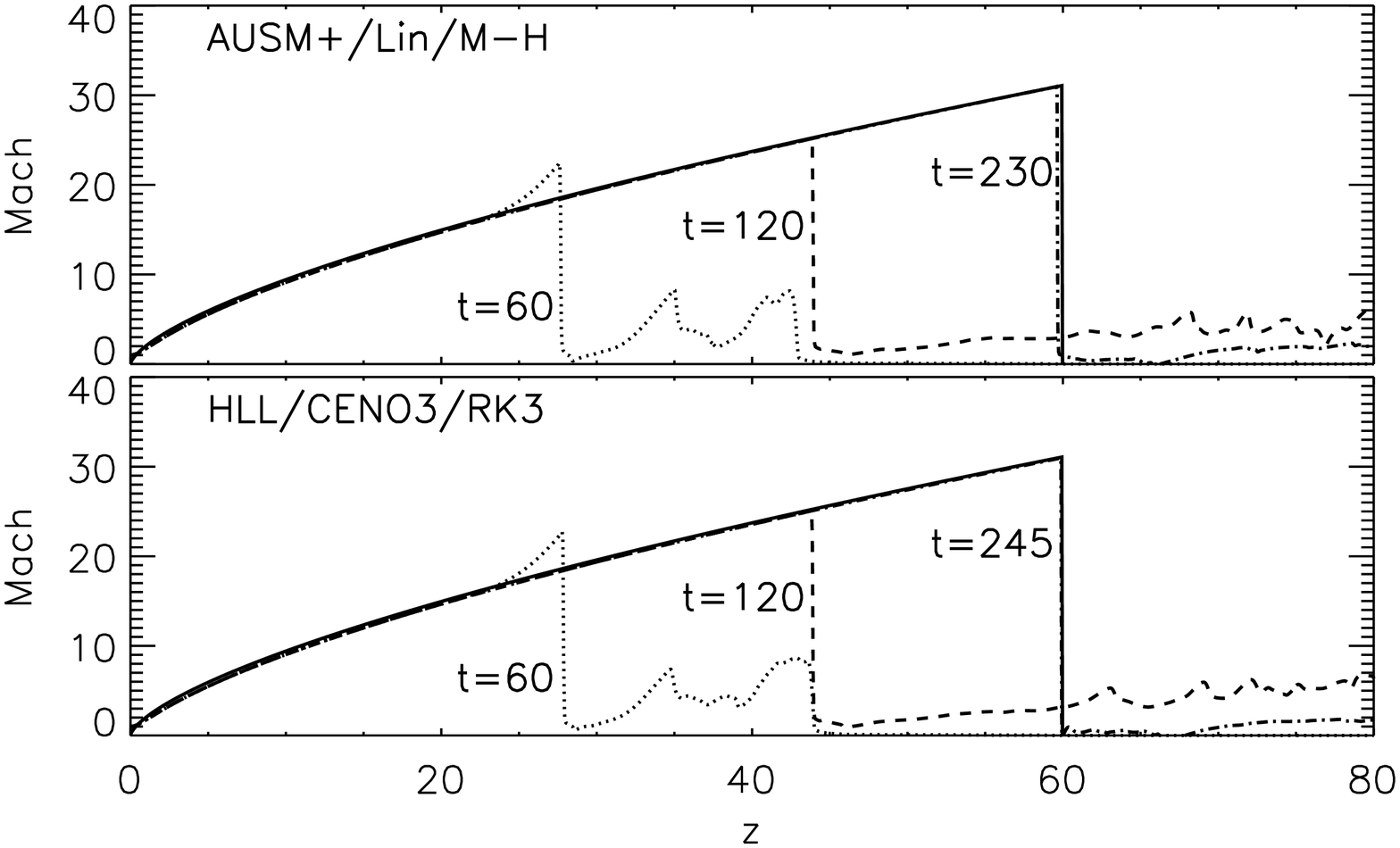}
 \caption{\footnotesize 
          Axial Mach numbers plotted at  
          $t=60,120,230$ (for Muscl Hancock) and 
          at $t=60,120,245$ (for RK3) using dotted, dashed and 
          dash-dotted lines, respectively. 
          Because of the small-amplitude oscillations around the 
          equilibrium position, the last time is not the same.
          The solid line gives the empirical relation (\ref{eq:knuth}) 
          for $0<z<z_\mathrm{m}$ where $z_\mathrm{m}$ is given by Eq. 
          (\ref{eq:young}).}
 \label{fig:hdjet_b}
\end{figure}

Code validation, in addition to the typical tests described above,
can be carried out against laboratory experiments of under-expanded free jets
as well. This kind of experiments consists of injecting, through a converging
nozzle, a gas at stagnation pressure $p_0$ into a chamber kept
at lower pressure $p_{\rm c}$. 
The shock structure that forms is typically axially symmetric and 
consists of a quasi-stationary Mach-disk, located at a distance 
$z_{\rm m}$ from the nozzle, and of a barrel and reflected shocks
\citep{Y75}.
After series of experiments with different gases, an
empirical expression was found \citep{AN59,BS61,AS66} 
that relates the Mach-disk location $z_{\rm m}$ 
with the pressure ratio \citep{Y75}:
\begin{equation}\label{eq:young}
z_{\rm m} = 1.34 \ r^* \sqrt{\frac{p_0}{p_{\rm c}}} \,,
\end{equation}
with $r^*$ the effective sonic nozzle radius. 
Furthermore, \cite{K64} obtained for the Mach number $M_{\rm axis}$ 
on the jet axis the following empirical expression:
\begin{equation}\label{eq:knuth}
M_{\mathrm{axis}} \approx (2.2)^{\frac{\Gamma-1}{2}} 
\Big[\Gamma(\Gamma-1) \Big]^{-\frac{\Gamma-1}{4}}
\left( \frac{\Gamma+1}{\Gamma-1} \right)^{\frac{\Gamma+1}{4}} 
\left( \frac{z}{2r^*} \right)^{\Gamma-1} \,.
\end{equation}

We have carried out 2D numerical simulations in cylindrical coordinates
on a uniform grid $r\in [0,40]$, $z\in[0,80]$ with $20$ zones per
beam radius. The grid has been further extended in both directions (up to $r=80$ and
$z=160$) by adding a second patch of geometrically stretched zones 
($100$ and $200$ zones in $r$ and $z$, respectively) in order to avoid spurious 
reflections at the boundaries.
Free outflow is set at the outer boundaries and 
reflective conditions are imposed on the axis $r=0$ and for $r>1$ at $z=0$.
Considering the actual nozzle as the injection zone at $z=0$, $r\le 1$, we have 
obtained the values of the pressure $p^*$ and 
density $\rho^*$ by employing the isentropic laws for
a perfect gas \citep{S83} for a converging nozzle with 
stagnation pressure $p_0$ and density $\rho_0$:
\begin{equation}\label{eq:isen}
    p^* = p_0 \left( \frac{2}{\Gamma+1} \right)^{\frac{\Gamma}{\Gamma-1}}  \,, \quad
 \rho^* = \rho_0 \left( \frac{2}{\Gamma+1}\right)^{\frac{1}{\Gamma-1}} \,.
\end{equation}
We choose $p_0/p_c = 2\cdot 10^3$ and $\rho_0/\rho_c = 2\cdot 10^4$ 
(typical of an argon jet in helium at ambient temperature) and 
accordingly choose a sonic injection velocity at the the nozzle ($M^* = 1$). 
Densities, velocities and lengths are normalized to the ambient $\rho$, 
sound speed and beam radius, respectively.

Fig. \ref{fig:hdjet_a} shows the density maps for the two selected 
numerical schemes (see Table \ref{tab:tests}) at $t=240$. 
The results obtained with the AUSM+ Riemann solver (scheme a)
disclose a somewhat greater amount of small scale structure than 
the HLL integration (scheme b). This is not surprising because of the
former's ability to properly capture contact and shear waves.
On the contrary, the latter greatly simplifies the wave pattern of the
Riemann fan at the cost of introducing extra numerical dissipation.
This is compensated, to some level, by the choice of the $3^{\rm rd}$ order 
interpolant (CENO3) which, nevertheless, is better combined with an equally accurate 
time marching scheme (RK3).
The overall balance, therefore, favors the second order scheme since only
$4$ Riemann problems must be solved instead of the $6$ required by RK3,
see Table \ref{tab:nriem}. This conclusion is supported not only by 
the decreased numerical viscosity inherent to scheme a), but also by the reduced 
computational cost, which sees scheme b) being $\sim 1.6$ times slower
than scheme a). 
  
In both cases we see that the Mach number profiles plotted in Fig \ref{fig:hdjet_b}
exhibit an excellent agreement with the empirical relation
(\ref{eq:knuth}).
It should be emphasized that the saturated position of the Mach disk 
oscillates around the experimental value, Eq. (\ref{eq:young}), 
by a few percents.  
The oscillations are caused by disturbances originated at the 
intersection between the Mach disk and the barrel shock surrounding
the jet.

%%%%%%%%%%%%%%%%%%%%%%%%%%%%%%%%%%%%%%%%%%%%%%%%%%%%%%%%%%
\subsection{Rotated MHD Shock Tube Problem}
\label{sec:mhd_st2.5}
%
%%%%%%%%%%%%%%%%%%%%%%%%%%%%%%%%%%%%%%%%%%%%%%%%%%%%%%%%%%

\begin{figure}
 \epsscale{1.0}
 \plotone{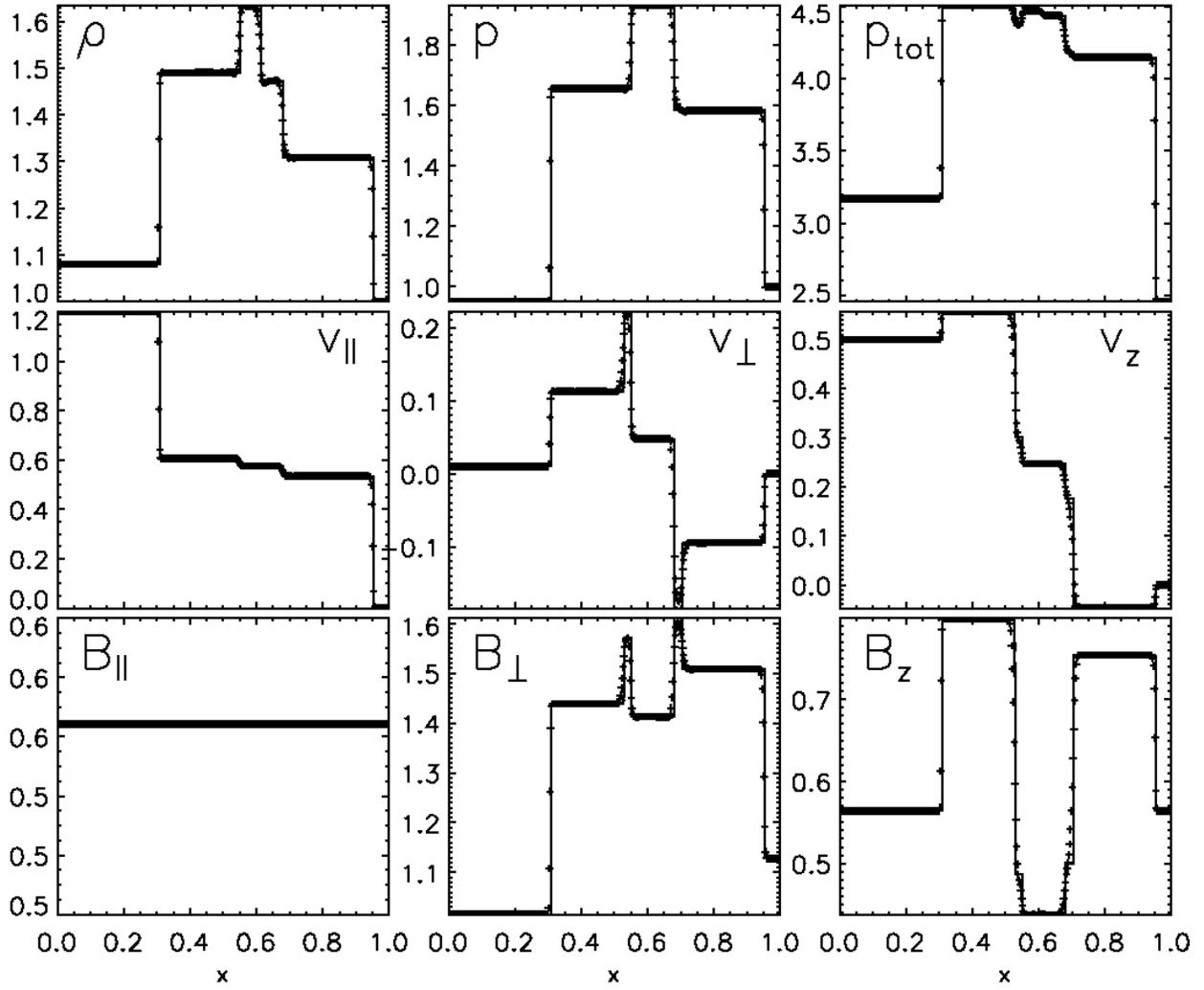}
 \caption{\footnotesize Rotated MHD shock-tube problem at $t=0.2\cos\alpha$. 
          Symbols give the solution computed on a two-dimensional grid
          with $400\times 2$ zone, 
          while the solid line gives a reference solution computed 
          for the non-rotated version at a resolution of $8192$ zones.}
 \label{fig:mhd_st2.5}
\end{figure}

\begin{figure}
 \epsscale{1.0}
 \plotone{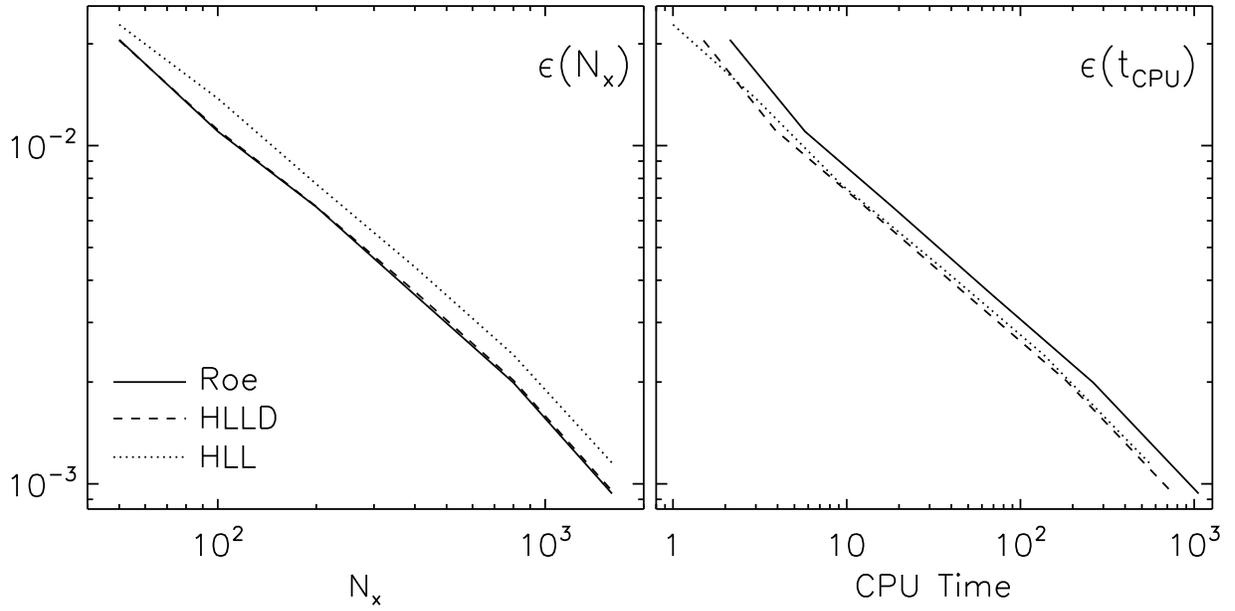}
 \caption{\footnotesize Errors in $L_1$ norm 
          ($\epsilon_{L_1}(q) \equiv \sum |q_{ij} - q^{\rm ex}_{ij}|\Delta x_i\Delta y_j$ 
           where $q_{ij}$ and $q^{\rm ex}_{ij}$ are the numerical and analytical
           solutions on the mesh)
          for the rotated MHD shock-tube problem at $t=0.2\cos\alpha$. 
          In order to consider all discontinuities 
          $\epsilon$ was computed as the arithmetic average of $\epsilon_{L_1}(\rho)$ 
          and $\epsilon_{L_1}(B_z)$.
          On the left, errors are plotted as function of the mesh
          size, whereas the left panel gives the errors as function of the 
          CPU time, normalized to the fastest integration.
          Solid, dashed and dotted lines refer to computations carried with the
          Roe, HLLD and HLL Riemann solvers, respectively.}
 \label{fig:mhd_st2.5_error}
\end{figure}

This test sets up a Riemann problem between left
and right states given, respectively, by
\footnotesize
\begin{equation}
 \left(\begin{array}{c} 
  \rho \\
   v_\parallel \\
   v_\perp  \\
   v_{\mathrm{z}} \\
   p_{\mathrm{g}} \\
   B_\parallel \\
   B_\perp \\
   B_{\mathrm{z}} \\
 \end{array}\right)_{\mathrm{L}}
    = 
 \left(\begin{array}{c} 
  1.08  \\
   1.2  \\
   0.01  \\
   0.5   \\
   0.95  \\
   2/\sqrt{4\pi}   \\
   3.6/\sqrt{4\pi}  \\
   2/\sqrt{4\pi}    \\
 \end{array}\right) 
\,,\quad
 \left(\begin{array}{c} 
  \rho \\
   v_\parallel \\
   v_\perp  \\
   v_{\mathrm{z}} \\
   p_{\mathrm{g}} \\
   B_\parallel \\
   B_\perp \\
   B_{\mathrm{z}} \\
 \end{array}\right)_{\mathrm{R}} 
    = 
 \left(\begin{array}{c} 
   1  \\
   0  \\
   0  \\
   0   \\
   1   \\
   2/\sqrt{4\pi}   \\
   4/\sqrt{4\pi}  \\
   2/\sqrt{4\pi}    \\
 \end{array}\right)\,, 
\end{equation}
\normalsize
where parallel ($\parallel$) and perpendicular ($\perp$) 
components refer to the direction of structure propagation.
The ratio of specific heats is $\Gamma = 5/3$.
The line $2y = (\HALF-x)$ (corresponding to a rotation angle of
$\alpha = \tan^{-1} 2$ with respect to the y-axis) is chosen as the 
surface of discontinuity separating the two constant states and 
the computational domain $[0,1]\times[0,1/N_{\mathrm{x}}]$ 
is discretized with $N_{\mathrm{x}} \times 2$ zones, as in \cite{Li05}. 
This setup yields a mesh spacing ratio $\Delta x/\Delta y = 2$
thus ensuring that the initial magnetic field is divergence-free
(CT is used to advect magnetic fields).
At the boundaries in the $y$ direction we impose translational 
invariance by applying shifted boundary condition, i.e., for
any pair of indexes $(i,j)$ spanning the ghost zones, we set
$V_{i,j} = V_{i-1,j+1}$ at the lower $y$-boundary 
and $V_{i,j} = V_{i+1,j-1}$ at the upper $y$-boundary.

The structure involves three-dimensional vector fields and the
outcoming wave pattern is bounded by two fast shocks 
(located $x\approx 0.3$ and $x\approx 0.95$) enclosing 
two rotational discontinuities (visible only in the magnetic field),
two slow shocks and one contact wave in at $x\approx 0.62$.
Fig. \ref{fig:mhd_st2.5} shows a cut along the $x$ axis  
at $t = 0.2\cos\alpha$ together with the high-resolution reference 
solution (solid line) computed from a one-dimensional integration at $t=0.2$.
The second order unsplit Runge-Kutta method with a CFL number of $0.4$
together with $2^{\rm nd}$ order linear interpolation on primitive
variabels and the Riemann solver of Roe were used for integration.

Computations were repeated at different resolutions starting from 
$N_{\mathrm{x}} = 50$ (lowest) up to $N_{\mathrm{x}} = 1600$ (highest), by 
doubling the number of zones each time. 
The performances and accuracies of the Roe, HLLD and HLL Riemann solvers 
were compared in terms of errors (in $L_1$ norm) and CPU time, as shown 
in Fig \ref{fig:mhd_st2.5_error}.
As indicated, HLLD and Roe yields comparable errors ($9.53\cdot 10^{-4}$ and
$9.36\cdot 10^{-4}$ at the highest resolution), whereas the HLL solver
results in errors which are $\sim 22\%$ larger. 
In terms of CPU time, the HLLD solver is considerably
faster than the Roe scheme and this is reflected in the right panel 
of Fig \ref{fig:mhd_st2.5_error}. Indeed, for a given 
accuracy, the computing time 
offered by HLLD significantly improves over the Roe solver and it is still
slightly better than the cheaper HLL scheme. In this sense, HLLD offers
the best trade off between accuracy and efficiency.

A caveat: the previous considerations hold, strictly speaking, for
the particular combinations of algorithms applied to the problem in question
and should not be trusted as general statements.
One may find, for example, that if interpolation is carried on characteristic
variables (instead of primitive) the overall trend which favours HLL over
Roe (in terms of CPU time) is reversed.
Still, our results illustrate how the choice of an algorithm or another
may considerably bear upon important issues of accuracy and efficiency.

%%%%%%%%%%%%%%%%%%%%%%%%%%%%%%%%%%%%%%%%%%%%%%%%%%%%%%%%%%%%%%%
\subsection{MHD Fast Rotor Problem}
\label{sec:mhd_rotor}
%
%
%
%%%%%%%%%%%%%%%%%%%%%%%%%%%%%%%%%%%%%%%%%%%%%%%%%%%%%%%%%%%%%%%

% ---------------------------------------------
%  * Geometry comparison
%  * Time stepping comparison: 
%    - for cartesian: ChTr (U), RK2(U), RK3(U) 
%    - for polar:               RK2(U), RK3(U)
%
%  Ribadire che CTU cartesian e' piu' 
%  conveniente perke' CFL e' piu' alto
% ---------------------------------------------

\begin{figure}
\epsscale{0.65}
\plotone{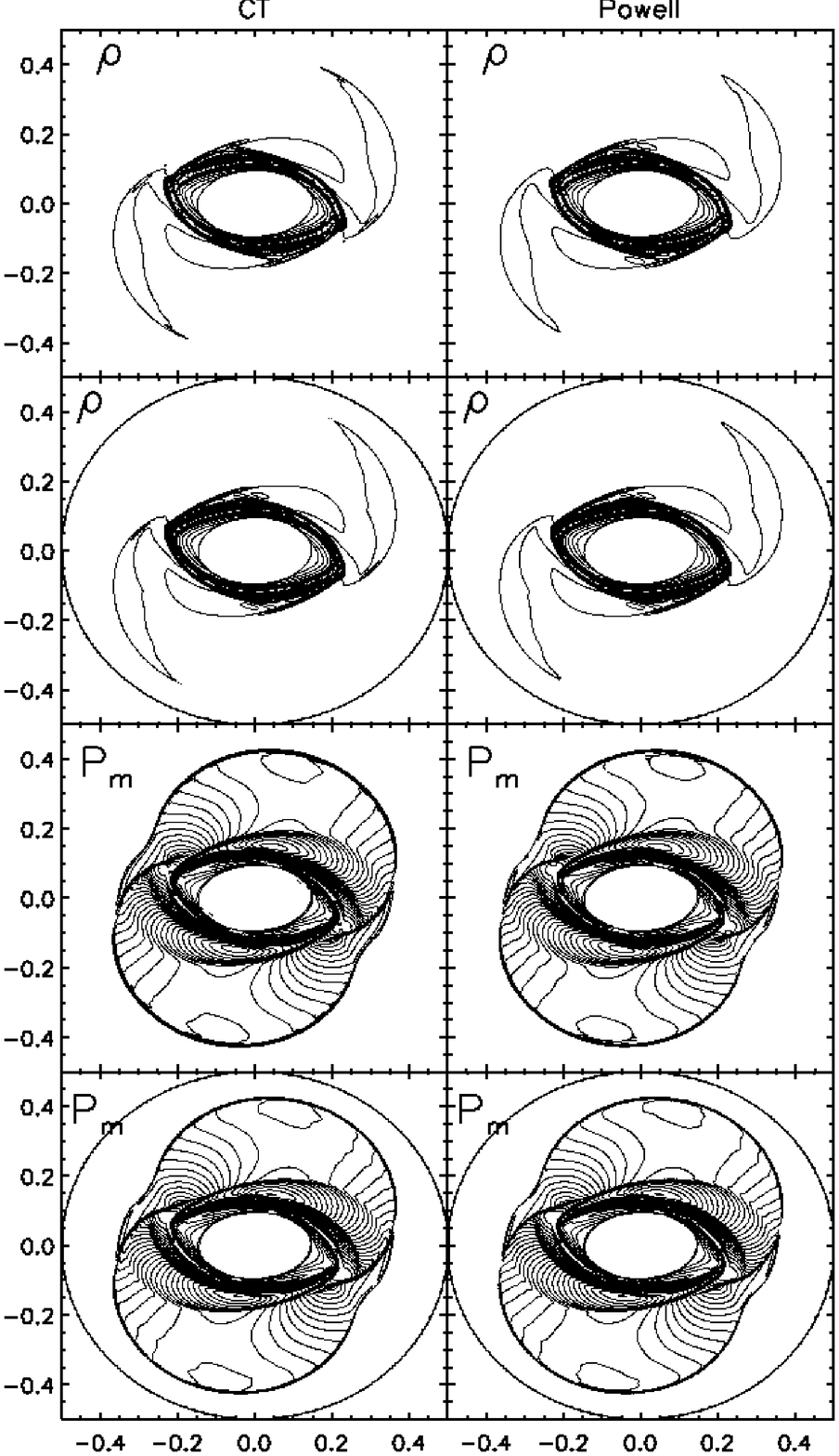}
 \caption{\footnotesize Contour maps for the MHD rotor problem at $t=0.15$. 
          Left and right panels show, respectively, computations carried with the CT and 
          Powell's method in Cartesian coordinates ($1^{\rm st}$ and $3^{\rm rd}$ rows from top) 
          and polar coordinates ($2^{\rm nd}$ and $4^{\rm th}$ rows from top).
          Thirty equally spaced contour levels are used for density 
          (($0.483 < \rho < 13.21$, $1^{\rm st}$ and $2^{\rm nd}$ rows) and magnetic 
          pressure ($0.0177 < |\vec{B}|^2/2 < 2.642$, $3^{\rm rd}$ and $4^{\rm th}$ ros).}
 \label{fig:mhd_rotor}
\end{figure}
A rapidly spinning ($\omega = 20$) cylinder with higher density is 
embedded in a static background medium with uniform pressure ($p = 1$),
threaded by a constant  magnetic field ($B_{\mathrm{x}} = 5/\sqrt{4\pi}$).
Hydrodynamical variables smoothly change their values between 
the disk and the external environment using a taper function:
\begin{equation}
(\rho, v_\phi) = \left\{\begin{array}{ll}
  (10, \omega r)      & \; \mathrm{for} \; r < r_0\,,   \\ \noalign{\medskip}
  (1+9f, f\omega r_0) & \; \mathrm{for} \; r_0\le r \le r_1\,, \\ \noalign{\medskip}
  (1, 0)              & \: \mathrm{otherwise} \,.
 \end{array}\right. 
\end{equation}
The disk has radius $r_0 = 0.1$ and the taper function 
$f = (r_1 - r)/(r_1 - r_0)$ terminates at $r_1 = 0.115$.
The initial radial velocity is zero everywhere and the
adiabatic index is $\Gamma = 1.4$.
This problem has been considered by several authors, see
for example \citet{BS99, LdZ04, Li05}, and references therein.

Because of the geometrical setting, we have setup the problem in both 
Cartesian and polar coordinates. 
On the Cartesian grid, the domain is the square $[-\HALF,\HALF]^2$ 
with outflow boundary conditions applied everywhere.
On the polar grid, we choose $0.05 < r < 0.5$, $0<\phi<2\pi$
with periodic boundary conditions in $\phi$ and zero-gradient
at the outer radial boundary. At the inner boundary we set 
$\partial_r (v_r/r) = \partial_r (v_\phi/r) = 0$.
For each geometry, we adopt the combination of algorithms listed in 
Table \ref{tab:tests} and compare two different strategies to control
the solenoidal constraint $\nabla\cdot\vec{B} = 0$, i.e., 
the CT and Powell's methods.

Fig. \ref{fig:mhd_rotor} shows density and magnetic pressure contours 
computed at $t=0.15$ on the Cartesian and polar grids at a resolution 
of $400^2$ zones.
As the disk rotates, strong torsional Alfv{\'e}n waves form and 
propagate outward carrying angular momentum from the disk 
to the ambient. Our results fully agree with those previously 
recovered by the above-mentioned authors and computations obtained with 
the CT and Powell's scheme shows excellent agreement.
Despite the higher complexity of the CT scheme where both staggered
and zone-centered magnetic field are evolved in time, the computational 
cost turned out to be comparable (CT/Powell $\sim 1.05$) for each 
system of coordinates. 
Moreover, the solutions computed in the two different geometries show nearly 
identical patterns, although polar coordinates (even with their intrinsically 
higher numerical viscosity at higher radii, due to the diverging nature of 
the grid) better fit the geometrical configuration of the problem.
However, computations carried out on the polar grid were
$\sim 5$ times slower than the Cartesian one, because of the more severe 
time step limitation at the inner boundary, where the grid narrows. 
These considerations can considerably affect the choice of geometry,
specially in long-lived simulations.

%%%%%%%%%%%%%%%%%%%%%%%%%%%%%%%%%%%%%%%
\subsection{Magnetized Accretion Torus}
\label{sec:mhd_torus}
%
%
%%%%%%%%%%%%%%%%%%%%%%%%%%%%%%%%%%%%%%%

% ---------------------------------------------
%  * Interpolation, R.S., Geometry: 
%    - HLLD Lin (Sph, Cyl)
%    - HLL Par  (Sph, Cyl) 
% ---------------------------------------------

\begin{figure}
\plotone{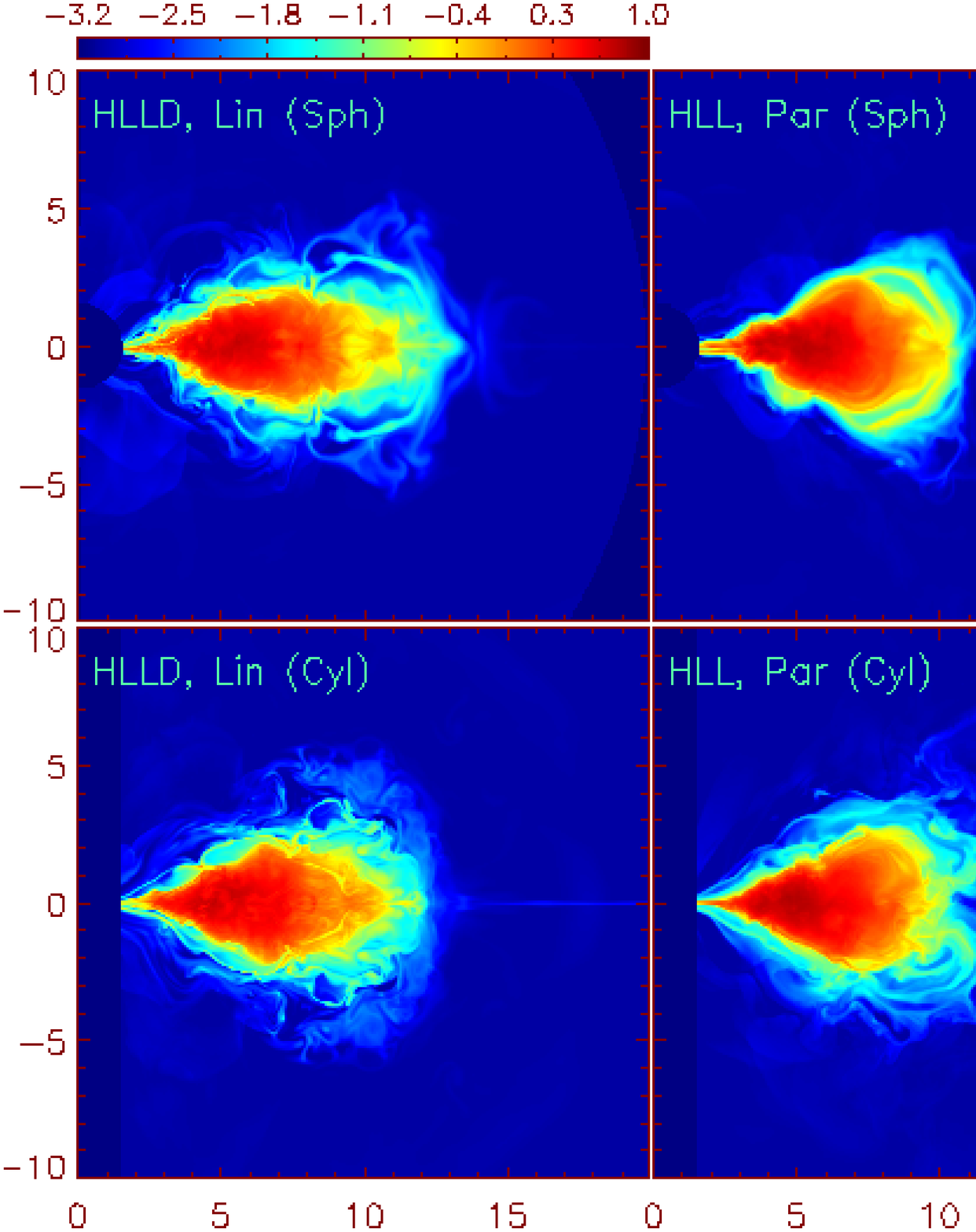}
 \caption{\footnotesize Density logarithm for the magnetized accretion torus at 
          $t/\Delta t = 5.6$. Computations with the HLLD Riemann solver (left) 
          and the HLL solver (right) are shown in spherical (top) and 
          cylindrical (bottom) geometries. The region $0<r<1.5$ is excluded 
          from computations.}
 \label{fig:mhd_torus}
\end{figure}

We now discuss an application of the code to a problem of 
astrophysical interest, along the lines of \cite{H00}.
The problem involves a magnetized, constant angular 
momentum ($\Omega \propto (r\sin\theta)^{-2}$) 
torus in a (properly normalized) pseudo-Newtonian 
gravitational potential, $\Phi = - (r - 1)^{-1}$. 
The torus has an equilibrium configuration described 
by the integral relation
\begin{equation}
 \frac{\Gamma p}{(\Gamma - 1)\rho} = C - \Phi - \frac{1}{2}
 \frac{l_{\mathrm{kep}}^2}{r^2\sin^2\theta}\,.
\end{equation}
Here $C$ is determined by the radial location of the inner edge of the torus 
($r = 3$ in our case) and $l_{\mathrm{kep}} = r^{3/2}/(r-1)$ is the Keplerian 
specific angular momentum at the pressure maximum ($r = 4.7$) where the orbital 
period is $\Delta T = 50$.
Density and pressure are tied by the polytropic relation $p = K\rho^\Gamma$,
with $\Gamma = 5/3$.
A poloidal magnetic field is initialized inside the torus using the $\phi$ component
of the potential vector $A_\phi \propto \min\left(\rho(r,\phi)  - 5, 0\right)$, 
normalized by the condition $\min\left(2p/|\vec{B}|^2\right) = 10^2$.

We solve the problem in spherical as well as cylindrical coordinates.
The spherical grid used for the problem is $1.5 \le r \le 20$ and $0 \le\theta\le \pi$,
with $320\times 256$ uniform zones covering the region $1.5 \le r \le 11.5$ and
$\pi \le 4\theta\le 3\pi$. The remaining regions were covered by a geometrically
stretched grid with $72$ zones. Outflow boundaries are imposed at the innermost and outermost
radii, whereas axisymmetry holds at $\theta = 0$ and $\theta = \pi$.
In cylindrical coordinates, the computational box is
$0 \le r \le 20$ and $-20 \le z \le 20$, with a uniform grid $320\times 320$
in the region $1.5 \le r \le 11.5$ and $-5 \le z \le 5$. Axisymmetry boundary
conditions are imposed at the axis, while all the rest are set to outflow.
In addition, each case is further investigated by adopting 1) the HLLD solver 
with second order slope limited reconstruction -case a and c in Table \ref{tab:tests}-
and 2) the HLL solver combined with piecewise parabolic reconstruction 
(case b and d in Table \ref{tab:tests}).
Second and third order Runge Kutta schemes are used for integration, respectively.

Figure \ref{fig:mhd_torus} shows the density logarithm at $t/\Delta T=5.6$
for the four cases respectively.
Initially, the strong shear generates a toroidal magnetic field component
leading to a pressure-driven expansion of the torus and accretion
takes place at $t/\Delta T \approx 1$. 
The growth of the poloidal field develops into a highly turbulent regime accounted for
by the magneto rotational instability \citep[MRI,][]{BH91} at $t/\Delta T\approx 4$. 
The effect is a net transport and redistribution of angular momentum.

Even though linear reconstruction presumably introduces more diffusion than 
parabolic interpolation, the HLLD solver, being more accurate, 
results in computations with a higher level of fine structure (compared to HLL) 
around the tori during the turbulent phase.
Besides, by normalizing the CPU time to case c), we found the ratio a:b and c:d 
to be $1:1.75$ (in cylindrical coordinates) and $1.55:2.42$ (in spherical 
coordinates), respectively.
Indeed, since higher than $2^{\rm nd}$ order interpolants are computationally more expensive 
and are used in conjunction with $3^{\rm rd}$ order time accurate schemes, 
the HLL integration comes at an extra computational cost.
This suggests that the use of a more accurate Riemann solver can balance the lower 
order of reconstruction in space and result, at the same time, in more efficient
computations.
Moreover, note that the equatorial symmetry breaks down more rapidly when parabolic 
reconstruction is employed, but it is retained to a higher level for the $2^{\rm nd}$ order
case.
This is likely due to the larger number of operations introduced by the PPM reconstruction 
step, inevitably leading to faster-growing round off errors.

Finally we note that computations in spherical geometry suffer from the major
drawbacks of introducing excessive numerical viscosity at the outer radii and
to severely limiting the time step at the inner boundary.
Both aspects are indeed confirmed in our numerical tests.
On the other hand, a cylindrical system of coordinate does not experience 
a similar behavior and thus a higher turbulence level is still observed 
at larger radii.

%Fig. \ref{fig:mhd_torus} shows the evolution at different orbital periods.
%Initially, the strong shear generates a toroidal magnetic field component
%leading to a pressure-driven expansion of the torus and accretion
%takes place at $t/\Delta T \approx 1$. 
%The growth of the poloidal field develops into a 
%highly turbulent regime accounted for by the magneto 
%rotational instability \citep[MRI,][]{BH91}, 
%$t/\Delta T=4$.
%The effect is a net transport and redistribution of angular momentum.

%This phase is followed by an evolution toward an almost Keplerian 
%angular momentum distribution and a quiescent state where the accretion 
%rate decrease ($t/\Delta T=10$), in agreement with \cite{H00}.

%%%%%%%%%%%%%%%%%%%%%%%%%%%%%%%%%%%%%%%%%%%%%%%%%%%%%%%%%%%%%%%%%%%
\subsection{Three-dimensional Spherically Symmetric Relativistic Shock Tube}
\label{sec:rhd_sph}
%
%
%
%
%%%%%%%%%%%%%%%%%%%%%%%%%%%%%%%%%%%%%%%%%%%%%%%%%%%%%%%%%%%%%%%%%%%

\begin{figure}
\epsscale{1.0}
\plotone{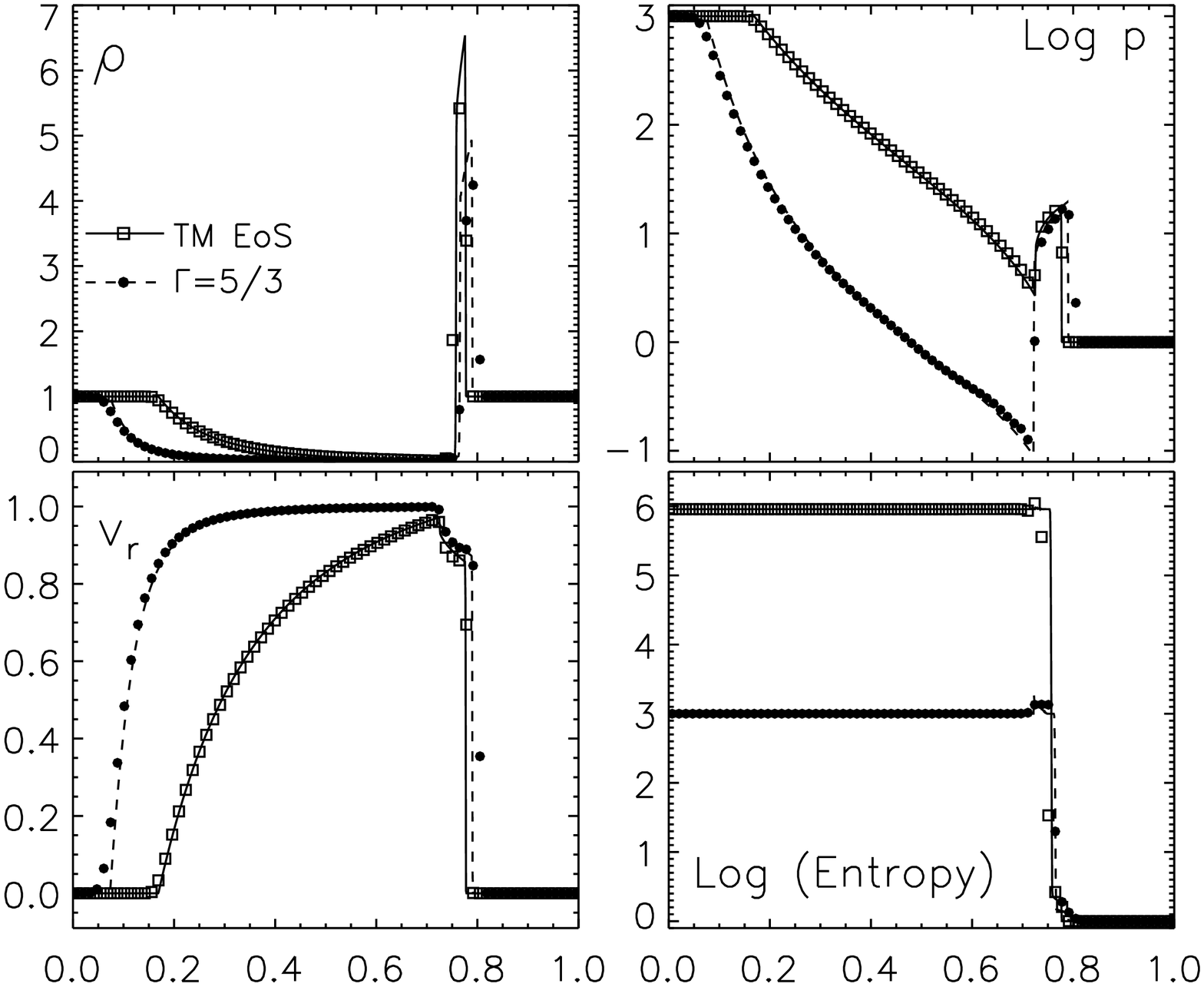}
 \caption{\footnotesize Spherically symmetric relativistic shock tube at $t=0.4$.  
          Solid and dashed lines give the one dimensional reference solutions for
          the TM and $\Gamma=5/3$ equations of state, respectively. 
          Density ($\rho$), pressure logarithm ($\log p$), radial velocity 
          ($v_r$) and entropy logarithm are shown.
          Over-plotted filled circles and empty boxes show the same quantities 
          along the main diagonal of the three-dimensional simulation.
          The CFL number is $0.8$ and are $128^3$ have been used.}
 \label{fig:rhd_sph}
\end{figure}
\begin{figure}
\epsscale{1.0}
\plotone{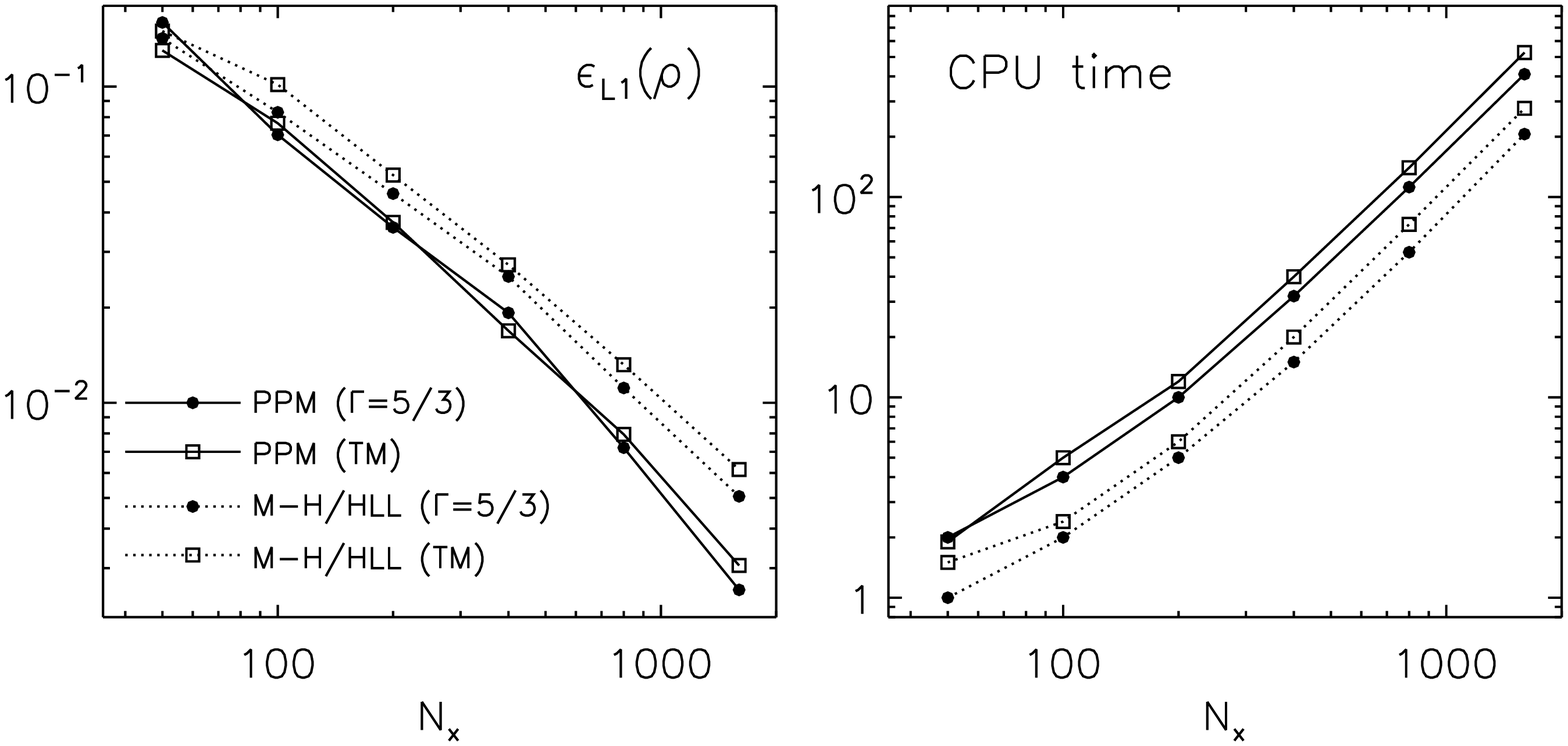}
 \caption{\footnotesize $L_1$ norm errors (on the left) and CPU time (on the right) for
          the one dimensional spherically symmetric relativistic shock tube as function
          of the resolution. 
          Solid and dotted lines label computations carried with the PPM scheme (a) and
          the MH scheme (b). Results pertaining to different equations of state
          are marked with filled circles ($\Gamma=5/3$) and boxes (TM EoS).
          The CPU time has been normalized to the fastest computation.}
 \label{fig:rhd_sph_error}
\end{figure}

An initial spherical discontinuity separates a 
high pressure region ($p=10^3$ for $r<0.4$) from a uniform medium
with $p=1$ (for $r>0.4$). The density is set to unity everywhere and the 
gas is initially static.

Computations proceed by investigating the influence of two different 
equations of state (EoS),
the constant $\Gamma$ law with $\Gamma=5/3$ (eq. \ref{eq:rhd_ideal}) and the 
TM equation (\ref{eq:rhd_tm}).
Moreover, we compare the performances of the PPM scheme (with the Two-Shock Riemann solver) 
with the simpler MH (with the HLL solver) which avoids characteristic decomposition 
of the equations (schemes a and b in Table \ref{tab:tests}). 
Spherical symmetry is assumed in the one-dimensional calculation.

The ensuing wave pattern (solid line in Fig \ref{fig:rhd_sph})
is comprised of a left rarefaction wave, a contact discontinuity in the middle 
and a right going shock. 
The overall morphology is significantly affected by the choice of the EoS.
Indeed, when the more realistic EoS is adopted, waves propagate at smaller 
velocities and this is particularly evident at the head and the tail points of the 
left-going rarefaction wave.
This results in a stronger adiabatic expansion for the constant $\Gamma$ gas
and in a denser shell between the shock and the trailing contact wave for the TM EoS.

Fig. \ref{fig:rhd_sph_error} reports the $L_1$ norm errors (for density) 
and the normalized CPU time as functions of the resolution 
$\Delta r^{-1}=50\cdot 2^n$ ($0\le n\le 5$) for both schemes and for the 
selected equations of state. Errors are computed against a reference solution 
of $16\cdot 10^3$ zones.
The cheaper scheme b) is certainly faster, although almost twice the resolution
should be employed to achieve comparable accuracy with scheme a).
The opposite trend is found for the PPM scheme which is slower by almost
a factor of $2$ when compared to MH/HLL.
Employment of the TM EoS results in an additional 
computational cost of about $25\div 30\%$ independently of the Riemann solver.
However, the simple results obtained here suggest that a more realistic EoS 
can significantly impact the solution and thus justify the direct 
use of Eq. (\ref{eq:rhd_tm}) instead of (\ref{eq:rhd_ideal}).
Further reading on this topic, together with its extension to RMHD, may 
be found in \cite{RCC06, MK07}.

Spherically symmetric problems can serve as useful verification benchmarks to 
tests the code performance on three dimensional Cartesian grids.
For this purpose, we use the second-order ChTr scheme with the recently developed
HLLC Riemann solver \citep{MB05} (scheme c) and adopt a resolution of $128^3$ 
computational zones.
By exploiting the symmetry, computations are carried out on one 
octant ($[0,1]^3$) only by adopting reflecting boundary conditions.
Results are over-plotted using symbols in Fig. \ref{fig:rhd_sph}. 
Small propagation structures such as the thin density shell pose serious
computational challenges. Indeed we see that the maximum density peaks
achieve $\sim 85\%$ and $\sim 83\%$ of the reference values for
$\Gamma=5/3$ and the TM equation of state, respectively. 
The smooth rarefaction wave is correctly captured at constant
entropy with very small overshoots at the back. 
Results at higher resolution (not shown here) show increasingly better 
agreement.

Scaling performances have been tested for this problem on different 
parallel platforms, all yielding satisfactory scaling properties.
Fig. \ref{fig:scaling} shows, for instance, the normalized execution 
time on the IBM SP Cluster 1600 p5-575 with $N_\mathrm{p} = 2^p$ 
processors ($1\le p \le 5$) for the same problem at 
a resolution of $128^3$, indicating an almost ideal 
scaling ($\propto 1/N_{\mathrm{p}}$).

%%%%%%%%%%%%%%%%%%%%%%%%%%%%%%%%%%%%%%%%%%%%%%%%%%%%%%%%%%%%%%%%%%%
\subsection{Three-dimensional Relativistic Magnetized Blast Wave}
\label{sec:rmhd_blast3d}
%
%
%
%
%%%%%%%%%%%%%%%%%%%%%%%%%%%%%%%%%%%%%%%%%%%%%%%%%%%%%%%%%%%%%%%%%%%

\begin{figure}
\epsscale{0.6}
\plotone{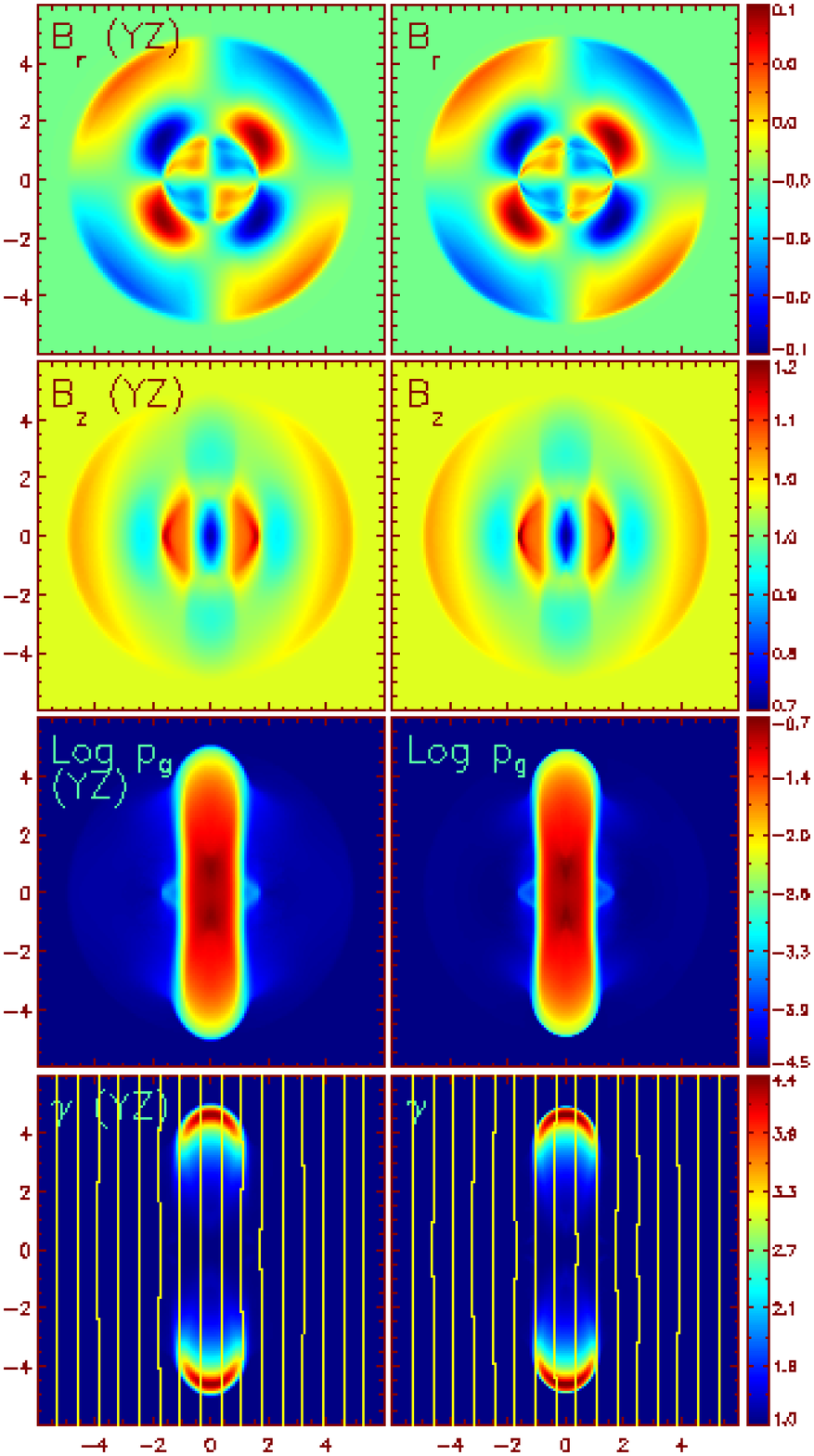}
 \caption{\footnotesize Relativistic magnetized blast wave at $t=0.4$ in Cartesian (left
          panels, cuts in the $y-z$ plane at $x=0$) and cylindrical 
          (right panels) coordinates.
          Shown in the uppermost panels are the cylindrical radial and 
          vertical magnetic field distributions with equally
          spaced color levels ranging from $-0.052$ to $0.052$ (for $B_r$) and
          from $0.74$ to $1.17$ (for $B_z$).
          The bottom panels show, respectively, thermal pressure (in log
          scale) and the Lorentz factor with contours ranging
          from $-4.54$ to $-0.74$ and from $1$ to $4.45$.
          Magnetic field lines are over-plotted on the Lorentz 
          factor distribution.}
 \label{fig:rmhd_blast3d}
\end{figure}

A spherical region with density $\rho = 10^{-2}$ and thermal 
pressure $p = 1$ is embedded in a static uniform medium with 
$\rho = 10^{-4}$ and $p = 3\cdot10^{-5}$.
The sphere is centered around the origin and has radius $r = 0.8$.
A linear smoothing function is applied for $0.8<r<1$.
The whole region is threaded by a constant vertical magnetic field in the 
$z$ direction, $B_{\mathrm{z}} = 1$.
The ideal equation of state with specific heat ratio $\Gamma = 4/3$ is used.
A similar setup in two dimensional planar geometry was earlier considered by
\cite{K99}.

Our computational grid employs $512^3$ zones on the box $[-6, 6]^3$.
By exploiting the symmetric nature of the problem we reduce the 
computations to one octant ($0<x,y,z<6$) at half the resolution
This is easily achieved by symmetrizing all quantities with respect 
to the coordinate planes and flipping the signs of
$(B_{\mathrm{x}}, v_{\mathrm{x}})$ at $y=0$, $(B_{\mathrm{y}}, v_{\mathrm{y}})$ at $x=0$,
$(v_{\mathrm{z}}, B_{\mathrm{x}}, B_{\mathrm{y}})$ at $z=0$.

Figs. \ref{fig:rmhd_blast3d} shows the magnetic field distribution, thermal
pressure and Lorentz factor for the two configurations at $t = 4$.
As a reference solution, computations on a cylindrical
axisymmetric grid with $[512\times1024]$ zones are also shown.
The expanding region is delimited by a fast forward shock 
propagating radially at almost the speed of light.
Because of the strong magnetic confinement a jet-like
structure develops in the $z$ direction with 
a maximum Lorentz factor of $\gamma_{\max}\approx 4.5$.
This problem is particularly challenging because of the very 
small plasma $\beta = 2p/|b_{\mathrm{m}}|^2 = 6\cdot 10^{-5}$
and instructive in checking the robustness of the code 
and the algorithm response to different kinds of degeneracies.

\begin{figure}
 \plotone{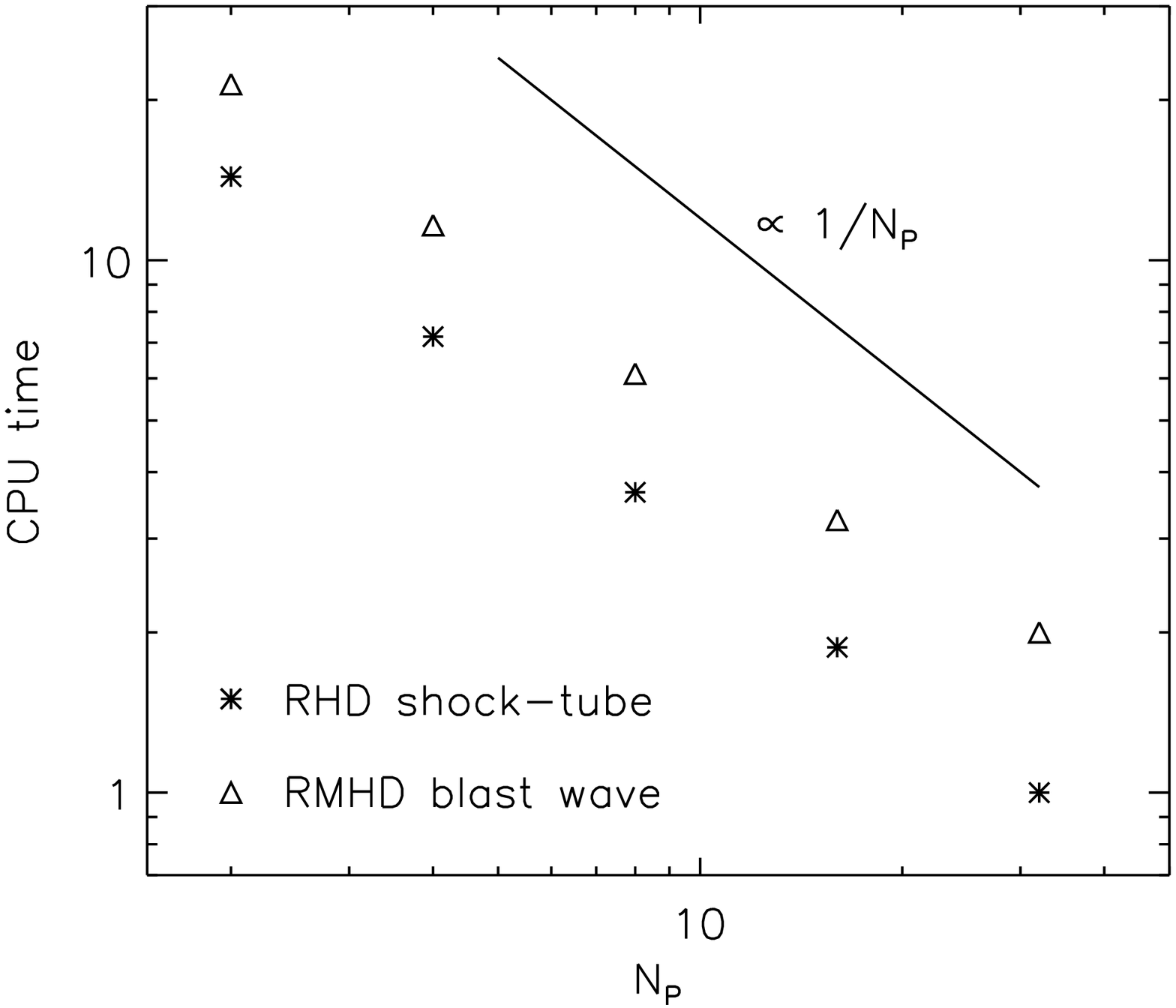}
 \caption{\footnotesize Scaling of PLUTO on 2, 4, 8, 16 and 32 processors 
          for the 3D relativistic shock tube interaction (stars) 
          and the RMHD blast wave (triangles) test problems, 
          respectively normalized to the 32- and 16- processor execution 
          time. A grid of $128^3$ was used. The perfect scaling slope
          ($\propto 1/N_{\mathrm{p}}$ where $N_{\mathrm{p}}$ is the
          number of processors) is shown as the solid line.  
          Computation were performed on the IBM SP Cluster 1600
          p5-575.}
 \label{fig:scaling}
\end{figure}

Scaling tests have been conducted in a way similar to 
\S\ref{sec:rhd_sph} and corresponding performance are shown in 
Fig. \ref{fig:scaling}.

%%%%%%%%%%%%%%%%%%%%%%%%%%%%%%%%%%%%%%%%%%%%%%%%%%%%%%%%%%%%%%%%%%%%%%%%%%%
\section{Conclusions}
%
%
%
%
%
%
%%%%%%%%%%%%%%%%%%%%%%%%%%%%%%%%%%%%%%%%%%%%%%%%%%%%%%%%%%%%%%%%%%%%%%%%%%%

In the context of astrophysical problems, where high Mach number 
compressible flow dynamics plays a crucial role,  
a sizable number of numerical codes is now available to the community
of scientists.
Most of them are based on the reconstruct-solve-average strategy 
typical of the high-resolution shock-capturing Godunov-type codes. 
The reasons for choosing one code or another can be the
most diverse, however, one can attempt to list a number of characteristics 
that a code should hold to be appealing for the skilled user, but not 
necessarily expert in numerics:
\begin{enumerate}
\item {\it a modular structure: the possibility to easily code and combine 
      different algorithms to treat different physics;}\\
      PLUTO offers a number of features which can be combined together.
      Besides four physical modules (e.g. Newtonian/relativistic hydrodynamics 
      with or without magnetic fields), 
      gravity, radiative cooling, resistivity, multidimensional 
      geometries, equations of state may be included 
      where required; 
\item {\it provide the user with a number of numerical schemes tested against 
      the most severe benchmarks;} \\
      several algorithms (e.g. Riemann solvers, interpolations, choice of 
      boundary conditions and so on) have been coded in PLUTO and their 
      choice is dictated by the problem at hand and/or by efficiency and 
      robustness issues.
\item {\it portability;} \\
      we have successfully ported PLUTO to a number of different
      Unix-based systems among which IBM sp5/sp4, SGI Irix, Linux Beowulf
      clusters, Power Macintosh. 
      In addition the code can run in either serial, single-processor
      or parallel machines.
\item {\it grid adaptivity: the ability to resolve flow features with 
      different spatial and temporal scales on the same computational
      domain;}\\
      the code provides a one-dimensional AMR integrator and
      multidimensional extensions are being developed by taking advantage 
      of the CHOMBO library (http://seesar.lbl.gov/ANAG/chombo/).
      This feature will be distributed with future versions of the code.
\item {\it last but not least, user friendliness;} \\
      a simple user-interface based on the Python scripting language 
      is available to setup a physical problem in a quick and 
      self-explanatory way. The interface is conceived to minimize
      the coding efforts left to the user.
\end{enumerate}

The code together with its documentation is freely distributed
and it is available at the web site http://plutocode.to.astro.it.

\acknowledgments
 We would like to thank S. Ritta for his useful contributions to the 
 under-expanded jet test and E. Beltritti for extensively 
 testing the code performances on several platforms.
 The present work was partially supported by the European Community 
  Marie Curie Actions - Human resource and mobility within the 
 JETSET network under contract MRTN-CT-2004 005592. 
 Numerical calculations were partly performed in CINECA (Bologna, Italy) 
 thanks to the support by INAF.

 A. M. would like to thank R. Rosner, T. Linde and T. Plewa and all 
 the people at the FLASH center at the University of Chicago for their 
 helpful suggestions and discussions during the early development of the code.

% =================================================================================

\end{document}